\documentclass[useAMS,usenatbib]{mn2e}

\usepackage{graphicx}
\usepackage{footmisc}
\usepackage{color}
\usepackage{rotating}
\usepackage{amssymb}
\usepackage{booktabs}
\title{Submm-bright X-ray absorbed QSOs at z$\sim$2: insights into the co-evolution of AGN and star-formation}
\author[Khan-Ali, et al.]{A. Khan-Ali\(^{1}\)\thanks{E-mail:
anuarkhan@ifca.unican.es}, F.J. Carrera\(^{1}\),
M.J. Page\(^{2}\), J.A. Stevens\(^{3}\), S. Mateos\(^{1}\), M. Symeonidis\(^{2}\)\(^{,4}\),
\newauthor J.M. Cao Orjales\(^{3}\)\\
\(^{1}\)Instituto de F\'\i{}sica de Cantabria (CSIC-UC), Avenida de los Castros s/n, Santander 39005, Spain.\\
\(^{2}\)Mullard Space Science Laboratory, University College London,
Holmbury St Mary, Dorking, Surrey RH5 6NT, UK.\\
\(^{3}\)Centre for Astrophysics Research, University of Hertfordshire, 
College Lane, Herts, AL10 9AB, UK.\\
\(^{4}\)University of Sussex, Department of Physics and Astronomy, Falmer, 
Brighton BN1 9QH, Sussex, UK.}
\begin{document}

\date{2014 March}

\pagerange{\pageref{firstpage}--\pageref{lastpage}} \pubyear{2014}

\maketitle

\label{firstpage}

\begin{abstract}
  We have assembled a sample of 5 X-ray-absorbed and submm-luminous
  type 1 QSOs at $z \sim 2$ which are simultaneously growing their
  central black holes through accretion and forming stars copiously.
  We present here the analysis of their rest-frame UV to submm
  Spectral Energy Distributions (SEDs), including new Herschel data.
  Both AGN (direct and reprocessed) and Star Formation (SF) emission
  are needed to model their SEDs.  From the SEDs and their UV-optical
  spectra we have estimated the masses of their black holes
  $M_{BH}\sim 10^{9}-10^{10}\,M_{\odot}$, their intrinsic AGN
  bolometric luminosities $L_{BOL}\sim(0.8 - 20)\times 10^{13}
  L_{\odot}$, Eddington ratios $L_{BOL}/L_{Edd}\sim 0.1 - 1.1$ and
  bolometric corrections $L_{BOL}/L_{X,2-10}\sim 30 - 500$. These
  values are common among optically and X-ray-selected type 1 QSOs
  (except for RX~J1249), except for the bolometric corrections, which
  are higher.  These objects show very high far-infrared luminosities
  $L_{FIR}\sim$ (2 - 8)$\times10^{12}\,M_{\odot}$ and Star Formation
  Rates SFR$\sim 1000 M_{\odot}/$y.  From their $L_{FIR}$ and the
  shape of their FIR-submm emission we have estimated star-forming
  dust masses of $M_{DUST}\sim 10^9\,M_\odot$.
  We have found evidence of a tentative correlation between the gas
  column densities of the ionized absorbers detected in X-ray
  (N$_{H_{ion}}$) and $SFR$.  Our computed black hole masses are
  amongst the most massive known. 
 
\end{abstract}


\section{Introduction}

In the last two decades, it has become clear that most local
spheroidal galaxy components (elliptical galaxies and the bulges of
spiral galaxies) contain a super massive black hole (SMBH) in their
centres. The proportionality between black hole (BH) and spheroid mass
suggests a direct link between the growth of the black hole as an
Active Galactic Nucleus (AGN) and the stellar mass of the spheroid
(e.g. \citealt{marconi04}, \citealt{kormendy}). Identifying the main
mechanisms for formation and evolution of galaxies, and their
interrelation to that of the growth of their central black holes is a
major issue in Astrophysics and Cosmology.

In the last decade, deep surveys at submillimetre (submm) and
millimetre (mm) wavelengths have identified a high-redshift population
of massive dusty galaxies that are undergoing extreme starbursts.
Since prodigious star formation (SF) is often obscured by dust, these
galaxies are luminous in the mm though far-infrared (FIR) wavebands
where the starlight absorbed by dust grains is re-emitted.  Likewise,
the hotter dust heated in the circumnuclear environment of an AGN will
emit at mid-infrared (MIR) wavelengths. Therefore, submm and MIR
observations can be combined to study activity in galaxies due to
dust-obscured starbursts and AGN. The launch of the \textit{Herschel Space Observatory}
data \citep{pilbratt} allow obtaining more accurate and
deeper measurements in the FIR, permitting better determinations of
the luminosity due to SF ($L_{SF}$) and hence the SFR.  In recent
years there have been many studies in which the SFR was compared with
the growth of the central BH (\citealt{rovilos}, \citealt{page13},
\citealt{rosario}, \citealt{lutz}) with conflicting results about the
relationship between AGN luminosity and $L_{SF}$ .

In this paper, we have studied a sample of X-ray-obscured QSOs,
described by \cite{page04} and \cite{stevens05}, and studied by
\cite{stevens04,stevens10}, \cite{page11} and \cite{carrera11} (20 unabsorbed objects), at
z$\sim$1-3 when most of the SF and BH growth are occurring in the
Universe. \cite{stevens05} found six detections at $> 5\sigma$ significance at 850$\mu$m (SCUBA). 
These QSOs have strong submm emission, much higher than typically found in QSOs at similar 
redshifts and luminosities. However for one of them (RX~J110431.75 + 355208.5) a synchrotron 
origin for the detected 850$\mu$m emission cannot be ruled out, 
so we have not followed up RX~J110431.75 in this work, and our sample consists of the other five sources.
They were specifically targeted with \textit{Herschel Space Observatory}
data OT 1 (PI: F.J. Carrera).

The ultraviolet (UV) and X-ray spectra of our QSOs show evidence for
strong ionized winds which produce the X-ray obscuration
\citep{page11}. Piecing all these clues together, we inferred that the
host galaxies of these QSOs are undergoing strong SF, while the
central SMBH are also growing through accretion. In principle the ionized winds are
strong enough to quench the SF \citep{page11} so, given that the QSOs have powerful submm emission, 
these objects are then caught at a special time, perhaps emerging from a
strongly obscured accretion state in an evolutionary stage which might
last about 10-15 per cent of the QSO lifetime \citep{hopkins}, which tallies nicely with X-ray absorbed 
broad line AGN being about 15 per cent of the X-ray broad line population (e.g. \citealt{page2000,mateos,corral,scott11,scott12}). 
Alternatively, the ionized winds are de-coupled from the SF, but then why only X-ray-absorbed 
AGN are submm-luminous remains to be explained. Some geometrical effect might be invoked, 
but it is difficult to see how to reconcile the dramatically different scales 
at which both processes are expected to happen.

Here, we endeavour to get the physical properties of the central QSOs
(luminosities, BH masses, Eddington ratios, etc.) and their host
galaxies (SFR, $M_{DUST}$, $M_{GAS}$, etc.) and their mutual
relationships (or lack thereof). In addition, we will try to fathom
their place in AGN-host galaxy co-evolution models.

The paper is organized as follows. In Section 2, we present our sample
and summarize the data used. In Section 3, we present the Spectral
Energy Distributions (SEDs) of all our objects, explain how we have
made fits to several representative templates and obtained results
from them. Moreover, we calculate the Black Hole properties from
UV-Optical spectra and we study the time scales of the evolution of
the AGN and the host galaxy. In section 4, we discuss all results and
compare them with those found in other samples.  Finally, we summarize
our findings in Section 5.

We have assumed throughout this paper a Hubble constant
$H_0=70$~km/s/Mpc, and density parameters $\Omega_\mathrm{ m}=0.3$ and
$\Omega_\Lambda=0.7$.

\section{Data}

We present in Table~\ref{herschelData} \textit{Herschel Space Observatory}
data for the fields around our QSOs (RX~J0057,
RX~J0941, RX~J1218, RX~J1249, RX~J1633).
These data include two photometric bands (100$\mu$m and 160$\mu$m)
from the PACS instrument \citep{poglitsch} and three photometric SPIRE
bands (250$\mu$m, 350$\mu$m and 500$\mu$m); \cite{griffin}, see Table
\ref{OthersData}.  The PACS observations were performed in two scan
orientations (70\textordmasculine and 110\textordmasculine) for each
band and field.  These images were combined using the standard
pipeline of Herschel Interactive Processing Environment (HIPE v8 and
v10).  The SPIRE were undertaken in smallmap mode. We used directly
Level 2 data obtained from the Herschel Archive and implemented a
script to remove the background from the data \citep{ukirt}. We also
present here new optical imaging data (R filter) from the William
Herschel Telescope for RX~J1249. The observations were taken on 20th
May of 2010. The reduction and calibration were carried out using
standard {\tt iraf} procedures (see \citealt{carrera11} for details).

In addition, part of the data used in this paper have been used or
obtained in other studies: X-ray spectroscopic observations from
XMM-Newton and optical-UV spectra from \citet{page01,page11}, Spitzer
and SCUBA photometric data from \citet{stevens04,stevens10}
and optical and infrared photometric data for RX~J0941 from
\citet{carrera11}.  Finally, we have also used public data from
different catalogues (SDSS, 2MASS, SUPERCOSMOS, WISE, GALEX and
OM-SUSS), see Table \ref{OthersData}.

We have used Sloan PSF magnitudes, converting them to fluxes directly using
$m_{AB}=-2.5\log(F_{\nu})-48.60$ with $f_{\nu_0}=3.63\times 10^3$Jy. For $u$ and $z$ we have converted previous to AB magnitude using \footnote{From
  \tt{http://www.sdss3.org/dr8/algorithms/fluxcal.php}}:

\begin{equation}
 u_{AB} = u_{SDSS}-0.04 \\
 z_{AB} = z_{SDSS}+0.02
 \label{magAB}
\end{equation}

For non-AB magnitudes we have used :

\begin{equation}
F_{\nu} = F_{\nu,0} \times 10^{-0.4 m}
\label{magnitudes}
\end{equation}

where $F_{\nu,0}$ are given in column 3 in Table~\ref{OthersData} for
each band.

A conservative 5 per cent error was added in
quadrature to the Herschel (extracted from HIPE documentation), SDSS and SUPERCOSMOS \citep{hambly} catalogued flux errors to account
for the uncertainties in the zero-points. In addition, we have also added in
quadrature a 2 per cent error for the 2MASS data \citep{cohen}.
Regarding data from WISE, we have added a 1.5 per cent uncertainty to the catalogued flux errors in all
bands to account for the overall systematic uncertainty from the
Vega spectrum in the flux zero-points. Additionally we have added a 10 per cent
uncertaintyto the 12 µm and 22 µm fluxes \citep{wright} to account the existing
discrepancy between the red and blue calibrators used for the
conversion from magnitudes to Janskys.

\begin{table*}
 \centering
 \caption{Summary of the Herschel observations on our fields.}
 \begin{minipage}{180mm}
  \begin{tabular}{ccrcccccc}
  \hline
  Object &  $RA$ &     $DEC$\hspace*{0.4cm}& z & OBSID PACS & Date Obs. & OBSID SPIRE & Date Obs. \\
 \hline
 RX~J0057 & $00:57:34.94$ & $-27:28:28.0$ & 2.19 &1342225345-6 & 2011-07-23 & 1342234705 & 2011-12-18 \\
 RX~J0941 & $09:41:44.61$ & $+38:54:39.1$ & 1.82 &1342232387-8 & 2011-11-17 & 1342246614 & 2012-06-03 \\
 RX~J1218 & $12:18:04.54$ & $+47:08:51.0$ & 1.74 &1342233430-1 & 2011-12-02 & 1342222665 & 2011-06-15 \\
 RX~J1249 & $12:49:13.85$ & $-05:59:19.4$ & 2.21 &1342235124-5 & 2011-12-24 & 1342224978 & 2011-07-31 \\
 RX~J1633 & $16:33:08.59$ & $+57:02:54.8$ & 2.80 &1342223963-4 & 2011-07-11 & 1342219634 & 2011-04-26 \\
\hline
\end{tabular}
  \label{herschelData}
\end{minipage}
\end{table*}

\begin{table*}
 \centering
 \caption{Photometric data used in this work. All the new data presented in this paper take into account systematic or calibration errors. These were added in quadrature to the catalogue errors.}
 \begin{minipage}{180mm}
  \begin{tabular}{cccccccccc}
  \hline
    Filter / Band & Wavelength & $F_{\nu,0}$ & RX~J0057 & RX~J0941 & RX~J1218 & RX~J1249 & RX~J1633 \\
     & ($\mu$m) & (Jy) & (mJy) & (mJy) & (mJy) & (mJy) & (mJy) \\
  \hline
    B$^a$ & 0.44 & 4270 & $0.125 \pm 0.007$ & -- & -- & -- & --  \\
    V$^*$ & 0.55 & 3670 & -- & -- & -- & $0.76 \pm 0.04$ & --  \\
    R$^b$ & 0.70 & 2840 & -- & $0.040 \pm 0.003$ & -- & -- & --  \\
    R$^f$ & 0.70 & 2840 & -- & -- & -- & $0.972 \pm 0.05$ & --  \\
    I$^a$ & 0.90 & 2250 & $0.152 \pm 0.009$ & -- & -- & -- & --  \\
    i$^b$ & 0.78 & -- & -- & $0.072 \pm 0.006$ & -- & -- & -- \\
    u$^{\prime c}$& 0.3551 & --  & -- & $0.0078 \pm 0.0010$ & $0.0130\pm 0.0012$ & $0.391 \pm 0.02$ & $0.011 \pm 0.0014$  \\
    g$^{\prime c}$& 0.4686 & --  & -- & $0.0175 \pm 0.0010$ & $0.0268 \pm 0.0015$ & $0.715 \pm 0.04$ & $0.0365 \pm 0.0019$ \\
    r$^{\prime c}$& 0.6165 & -- & -- & $ 0.038\pm 0.007$ & $0.035\pm 0.002$ & $1.02 \pm 0.05$ & $0.0448 \pm 0.0008$ \\
    i$^{\prime c}$& 0.7481 & -- & -- & $0.071 \pm 0.004$ & $0.043 \pm 0.003$ & $1.13 \pm 0.06$ & $0.045 \pm 0.002$ \\
    z$^{\prime c}$& 0.8931 & --  & -- & $0.101 \pm 0.007$ & $0.043 \pm 0.004$ & $1.37 \pm 0.08$ & $0.052 \pm 0.004$ \\
    J$^b$ & 1.25 & 1650 & -- & $0.14 \pm 0.02$ & -- & -- & -- \\
    K$^b$ & 2.2 & 673 & -- & $0.175 \pm 0.018$ & -- & -- & -- \\
J$^d$ & 1.235 & 1594 & $0.18 \pm 0.05$ & -- & -- & $1.55 \pm 0.07$ & -- \\
H$^d$ & 1.662 & 1024 & $0.36 \pm 0.07$ & -- & -- & $1.96 \pm 0.07$ & -- \\
K$_s^d$ & 2.159 & 667.7 & $0.30 \pm 0.07$ & -- & -- & $2.54 \pm 0.11$ & -- \\
WISE1$^g$ & 3.5 & 309.540 & $0.306 \pm 0.014$ & $0.311 \pm 0.012$ & $0.338 \pm 0.012$ & $2.06 \pm 0.06$ & $0.086 \pm 0.004$ \\
WISE2$^g$ & 4.60 & 171.787 & $0.55 \pm 0.02$ & $0.68 \pm 0.02$ & $0.328 \pm 0.016$ & $3.75 \pm 0.10$ & $0.101 \pm 0.006$ \\
WISE3$^g$ & 11.56 & 31.674 & $2.43 \pm 0.3$ & $2.4 \pm 0.3$ & $1.06 \pm 0.16$ & $17 \pm 2$ & $0.23 \pm 0.05$ \\
WISE4$^g$ & 22.09 & 8.363 & $4.4 \pm 1.0$ & $8.4 \pm 1.3$ & $4.4 \pm 1.0$ & $34 \pm 4$ & $0.8 \pm 0.3$ \\
IRAC Ch2 & 4.5 & -- & $ 0.476\pm 0.015$ & $0.713 \pm 0.019$& $0.149 \pm 0.009$ & $3.67 \pm 0.04$ & $0.083 \pm 0.006$ \\
IRAC Ch4 & 8 & -- & $1.53 \pm 0.03$ & $2.13 \pm 0.04$ & $0.363 \pm 0.015$ & $12.76 \pm 0.09$ & $0.153 \pm 0.010$ \\
MIPS & 24 & -- & $4.11 \pm 0.11$ & $5.25 \pm 0.18$ & $3.28 \pm 0.15$ & $26.31 \pm 0.13$ & $0.71 \pm 0.04$ \\
PACS100 & 100 & -- & $18 \pm 8$ & $21 \pm 7$ & $11 \pm 5$ & $89 \pm 10$ & $8 \pm 5$ \\
PACS160 & 160 & -- & $31 \pm 10$ & $39 \pm 10$ & $34 \pm 11$ & $98 \pm 10$ & $9 \pm 6$ \\
SPIRE250 & 250 & -- & $60 \pm 7$ & $89 \pm 5$ & $84 \pm 5$ & $111 \pm 6$ & $14 \pm 3$ \\
SPIRE350 & 350 & -- & $44 \pm 4$ & $90 \pm 5$ & $71 \pm 5$ & $72 \pm 5$ & $11 \pm 3$ \\
SPIRE500 & 500 & -- & $25 \pm 4$ & $68 \pm 5$ & $47 \pm 5$ & $32 \pm 4$ & $11 \pm 4$ \\
SCUBA450 & 450 & -- & $32 \pm 10$ & $45 \pm 10$ & -- & $35 \pm 8$ & -- \\
SCUBA850 & 850 & -- & $7.8 \pm 1.0$ & $12.4 \pm 1.0$ & -- & $11.0 \pm 0.7$ & $6.9 \pm 0.7$ \\
  \hline
\end{tabular}
  \label{OthersData}
\end{minipage}
\begin{minipage}{180mm}
\noindent $^a$ SUPERCOSMOS.\\
\noindent $^b$ \citealt{carrera11}.\\
\noindent $^c$ SDSS DR7. \\
\noindent $^d$ 2MASS. \\
\noindent $^f$ William Herschel Telescope. \\
\noindent $^*$ NED. \\
\noindent $^g$ WISE All-Sky. \\
\end{minipage}
\end{table*}

\section{Results}

\begin{figure*}
  \centering
   \includegraphics[width=1\textwidth]{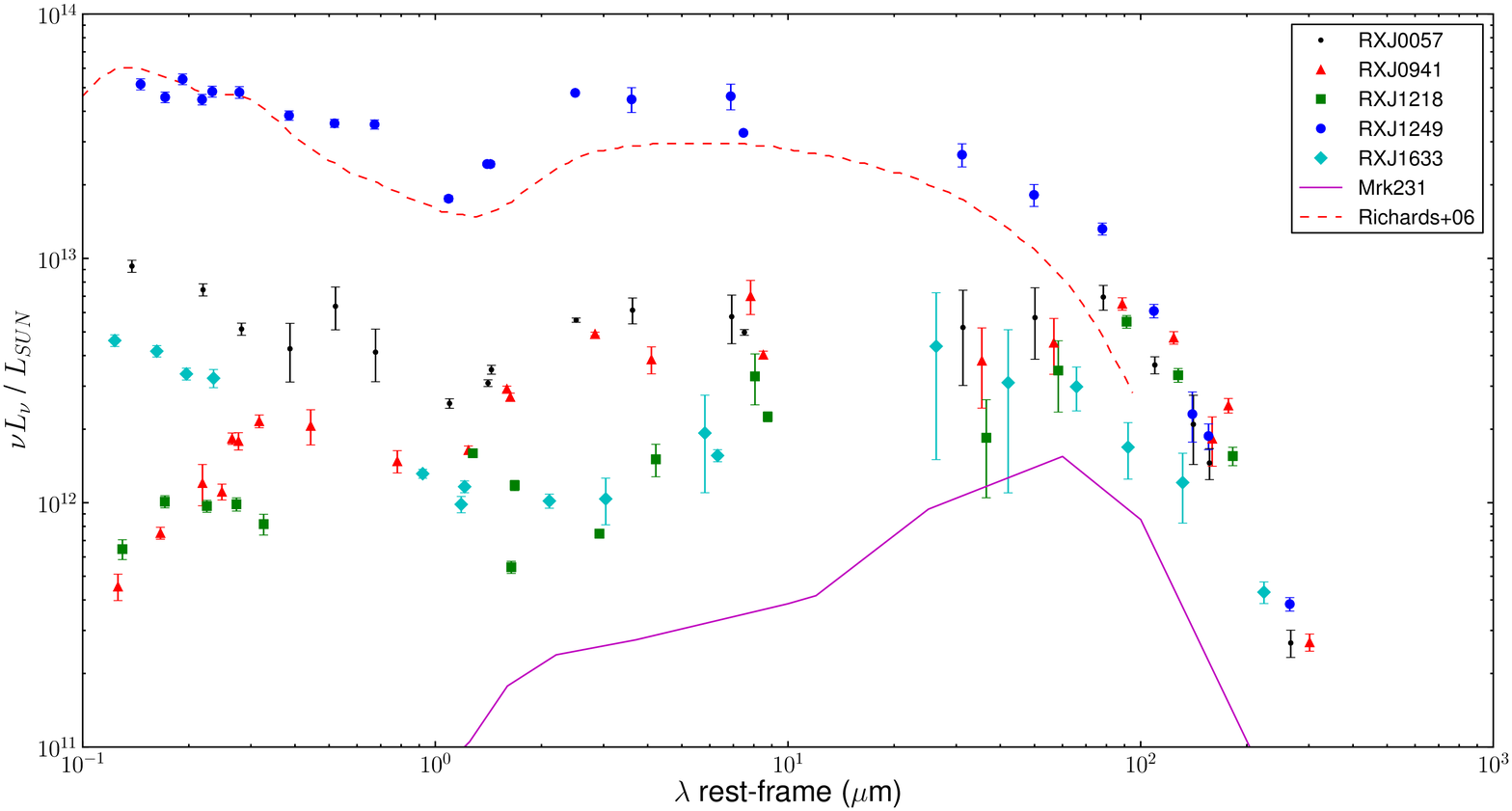}
   \caption{SEDs of our five QSOs, compared with a standard
     QSO template (\citealt{richards}, dashed line, normalized to our
     most luminous object at $\sim$2000~\AA) and the observed SED of
     Mrk~231 (a prototype ULIRG containing an AGN).} 
   \label{figAllSEDs}
\end{figure*}

We have constructed the spectral energy distributions of all
our objects, in $\nu L_{\nu}$ versus rest-frame wavelength in $\mu$m
(Figure \ref{figAllSEDs}).  We also show the observed SED (from NED)
of Mrk~231: an ultraluminous infrared galaxy (ULIRG), without any
renormalization. The average type 1 QSO template of \cite{richards} is
shown too. They constructed the SED from 259 quasars with SDSS and
Spitzer photometry, supplemented by near-IR, GALEX, VLA and ROSAT
data, where available.  We have rescaled this template to be close to
the observed SED of our brightest object (RX~J1249), whose UV-FIR
shape it closely matches.

All objects clearly show at least two components: UV-NIR contribution
attributable to direct accretion disk emission (as expected from their
type 1 nature), intrinsically absorbed in the cases of RX~J0941 and
RX~J1218; and a reprocessed thermal component in the MIR region from
warm optically thick dust further away from the nucleus (the torus).
In all of them we also observe an additional FIR/submm component
associated to cooler dust, heated by star formation (SF). We thus
confirm the presence of strong FIR emission due to SF in these
objects, at the ULIRG/HLIRG level (compared to e.g. Mrk 231).

\subsection{Fit models}
\label{fits}

Our next goal is to make fits to our data with different templates to
extract quantitative information about our objects. 
We use relatively simple empirical and theoretical templates, 
aiming at reproducing the general shape of the SEDs of our objects. 
We do not attempt to extract detailed physical information about our objects 
from those templates, since our data do not warrant such undertaking, and 
the underlying physics is likely to be more complex 
than that considered in the models. 
We have selected
the \textit{Sherpa: CIAO's modeling} \& \textit{fitting package}
\citep{sherpa} module for the Python platform \citep{sherpaPy} to
perform the fitting.  SEDs were fitted in $\nu L_{\nu}$ versus
rest-frame $\lambda$.
We have fitted only rest-frame wavelengths longer than 1218~\AA, to
avoid issues with the Ly-$\alpha$ forest.  We have corrected for
extinction in our Galaxy using the $A(\lambda)/A(V)$ values from Table
21.6 in \cite{cox} in the observer frame, assuming $R_V=3.1$ and
$A(V)/N(H)\sim5.3\times 10^{-22}$~cm$^{-2}$, extrapolating for
wavelengths longer than 250~$\mu$m assuming $A(\lambda)\propto
\lambda^{-2}$.

We have modelled the distinct constituents that can be appreciated in
the SEDs of our objects (see above and Figure~\ref{figAllSEDs}) with
three different components:

\begin{itemize}

\item An AGN accretion disk component: This template models the direct emission 
from the accretion disk of the AGN (sub-parsec scales). We use the pure disk newAGN4 template from
  \cite{rowan}, affected by intrinsic extinction (see below). The only
  free parameter is the normalization of the template. We have
  normalized the template to its integral in the 0.12-100~$\mu$m range
  (standard limits used to estimate the accretion disk luminosity).
  From the fits we directly obtain the value of the disk luminosity in that range,
  $L_{DISK}$.

\item A torus component: This template models the re-emission from the warm and hot dust 
 (on tens of parsecs scales, beyond the sublimation radius, see \citealt{antonucci}) that is warmed by the accretion disk emission. We have used both an empirical
  template from \cite{rowan} (dusttor, based on an average quasar spectrum) and three dusty clumpy torus models
  from \cite{nenkova} found by \cite{roseboom} to represent the
  average properties of type 1 QSOs (torus1, torus2 and torus3).
  The templates from \cite{nenkova} have $N_0 \sim 5-15$
  dusty clouds along radial equatorial rays. The angular 
  distribution of the clouds must have a soft edge, and the radial distribution decreases as $1/r$ or $1/r^2$. 
  
  The difference of the latter model with the
  others is fundamentally that it has a higher inclination of the torus
 with respect to the line of sight (20~deg versus 0~deg), a larger
 number of absorbing clouds (by a factor of three) and a different
 distribution of the clouds (constant versus declining with distance as
 $1/r$).
  
  Similarly, we have normalized them to their integral in the
  1-300~$\mu$m range (again, standard limits used for the torus
  luminosity), so the only free parameter is $L_{TORUS}$.

\item A SF component: we have used a subset of the
  \cite{siebenmorgenSB} spherical smooth models, found by \cite{myrto}
  to encompass the observed SEDs of star-forming galaxies at least up
  to $z\sim 2$. The full \cite{siebenmorgenSB} models have five free
  parameters to obtain their 7000 templates. The subset of templates
  recommended by \cite{myrto} (around 2000 templates) are divided into
  11 sub-grids, depending on the maximum radius and different star
  populations. For each source, we found the best-fit template for
  each of these 11 sub-grids.  We did not attempt to extract detailed
  physical information from the particular best-fit templates, since
  instead we were looking for physically-motivated templates that
  reproduced the spectral shape of the data. Therefore, the only
  parameter that we obtained from these fits, apart from the best-fit
  templates, was the integrated FIR luminosity (40-500$\mu$m)
  $L_{FIR}$, since all templates were normalized in that spectral
  range. In addition, we have found the optically-thin greybody (GB) model (with free
  temperature $T$ and slope $\beta$) that best matches each best-fit
  SF templates at wavelengths longer than 40~$\mu$m rest-frame,
  providing two more ``parameters'' associated to each fit: $T$ and
  $\beta$.

\end{itemize}

The direct AGN accretion disk component was affected by intrinsic
extinction modelled in terms of the hydrogen column density $N_H$, so
that absorption $\propto \exp(-\sigma(\lambda)N_H)$, where

\begin{equation}
\sigma(\lambda)={\log_e 10 \over 2.5}{A_V \over N_H}{A(\lambda) \over A(V)}
\end{equation}

We have used $A_V/N_H=0.76\times 10^{-22}$ and the $A(\lambda)/A(V)$
values from the \cite{gordon} \footnote{We obtained the numerical
  values from {\tt
    http://www.stsci.edu/$\sim$kgordon/\\Dust/Extinction/MC\_Ext/mc\_ave\_ext.html}}
SMC results for $\lambda<8100$~\AA, and from \cite{cox} for longer
wavelengths, both parametrizations merging smoothly at that
wavelength.

In summary, we have 4 free parameters in the fits: the three
luminosities ($L_{DISK}$, $L_{TORUS}$ and $L_{FIR}$) and the intrinsic
extinction ($N_H$). In total, we have 44 (4 torus models $\times$11 SF
models) possible combinations of all possible components and models.

We have minimized the usual $\chi^2$ statistics to choose between
different models. Attending to the fact that there are many ``hidden''
parameters in our modelling (the different torus templates, and the
many parameters in the \citealt{nenkova} and \citealt{siebenmorgenSB}
models) and to the bad reduced $\chi^2$ values of our best fits
($2-135$), we have neither attempted to use the parameters of the absolute
best-fitting template as the best value of the parameters, nor used
$\Delta\chi^2$ to estimate the uncertainties in those. Instead, 
we have taken into account both the dispersion of the parameter values among models and the relative errors on the photometric points.

We
have inspected the distribution of $\chi^2$ values for each source,
finding that there were ``families'' of models that produced quite
different values of that statistics. For each source (see below) we
have chosen the family of models with the lowest $\chi^2$.
Table~\ref{LumResults} shows the average values of the luminosities
and $N_H$ values among each best-fit family for each source and the minimum
value of $\chi^2$ for each best-fit family.
We have estimated the uncertainties in those values from the standard
deviation. 

Since have only included among those ``families'' one model for
  the direct AGN emission and only one or two models for the
  reprocessed AGN emission, the above dispersion gives rise to a
  unreasonably low uncertainty in the corresponding luminosities.
  Hence, we have used the average relative error on the relevant
  photometric points to estimate an additional relative uncertainty in
  all luminosities.  We have calculated the average relative error in
  the rest-frame ranges 1218-10000~\AA, 3-30~$\mu$m and $>100\,\mu$m
  for the direct AGN, reprocessed AGN and SF components, respectively.
  Those average relative values were multiplied by the
  corresponding average luminosities, finally adding them in quadrature to
  the above standard deviation.

  We believe that these values and their uncertainties are fair
  estimates of the luminosities of each component and of the effects
  of the photometric errors and our lack of an accurate physical model
  for what is really happening in each source.

The shape of most of our SEDs in Figure~\ref{figAllSEDs} is very
similar to the \cite{richards} (R06) QSO template (UV-MIR region), but with an excess in the
region of star formation (far-IR $/$submm region). So we have repeated
the fits replacing the obscured disk and torus components, with intrinsic extinction, 
type 1 QSO SED template from \cite{richards}, keeping the SF
component (a total of 11 sub-grids per fit). We have obtained $\chi^2$
values and FIR luminosities which are quite similar and compatible
with the previous three-component ones. From these results 
we can draw two main ideas:
\begin{itemize}
 \item The shape of our objects is well modelled (except perhaps for RX~J1633, see below) 
 by the \cite{richards} template (UV-MIR region) and SF component, with some intrinsic extinction in two cases: 
 they do not appear to be unusual in the UV-MIR region. Fits with separate disk and torus 
 components are similarly good. In what follows we quote the latter results because 
 they allow a more straightforward estimation of the direct and reprocessed AGN emission, 
 and because they are clearly better for RX~J1633.
 \item There is an excess in the far-IR $/$submm region that does not look 
 like emission from the torus, indeed it is fitted very well with SF templates. We obtain 
 quite similar FIR luminosities using accretion disk and torus models together or \cite{richards} template.
\end{itemize}

On the other hand, all of our objects (except RX~J1633) prefer the torus models dusttor, torus1 and torus2 
which have similar shape. However RX~J1633 clearly prefers the torus3 model. This torus model is
one of the most extreme from \cite{nenkova} with respect to the cold gas emission: it has a peak at higher wavelengths, 
but it is not enough to model our excess in far-IR $/$submm region. Summarizing, this (together with the discussion
at the beginning of the Section \ref{SFRsection}) reinforces our idea that the excess 
far-IR $/$submm region corresponds to SF emission and not to the reprocessed AGN emission.

In the following, we have used
the full three component fits for all QSOs because they treat the
direct and indirect AGN components separately, allowing a comparison
between them.

\begin{figure*}
  \centering
   \includegraphics[width=0.49\textwidth]{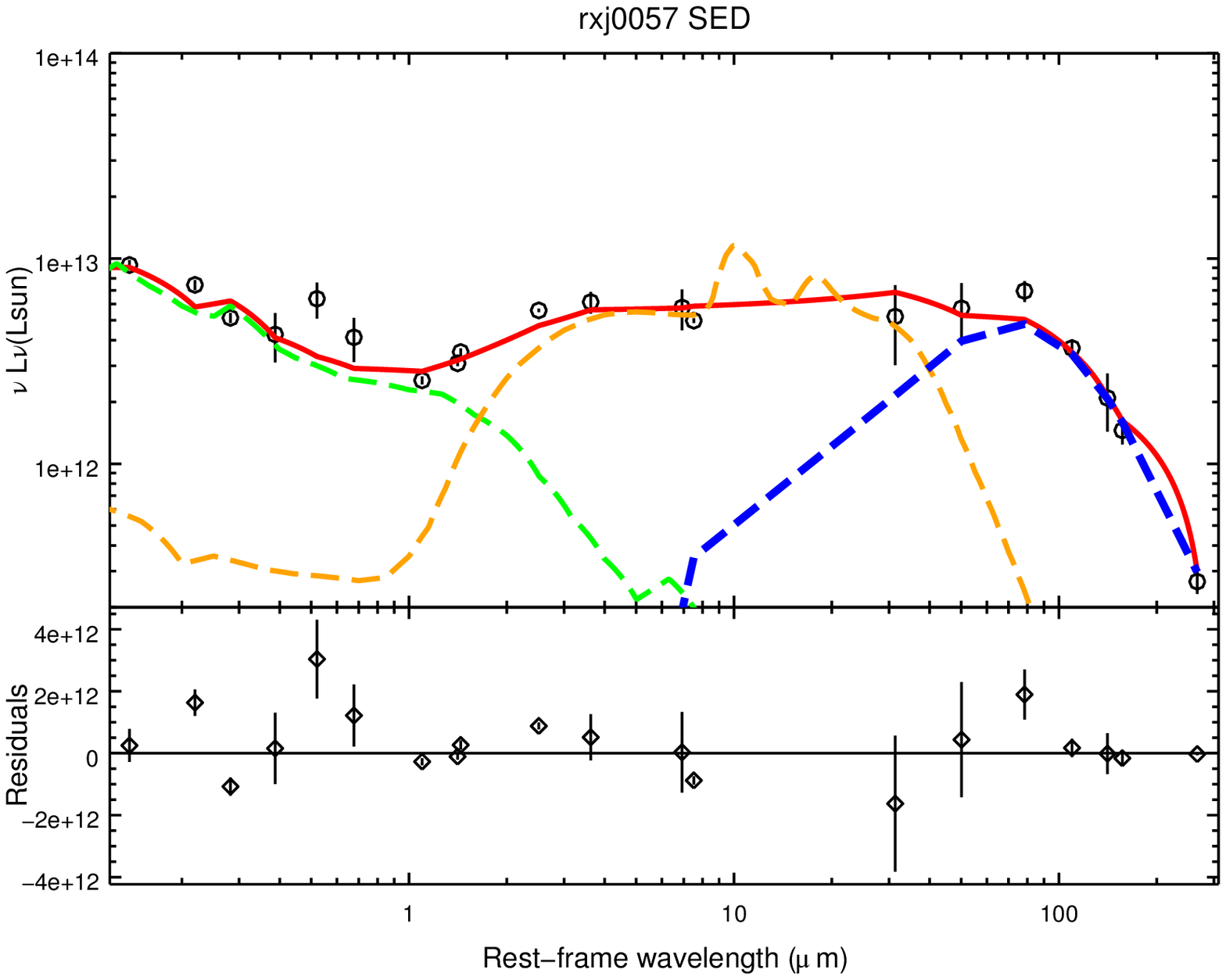}
   \includegraphics[width=0.49\textwidth]{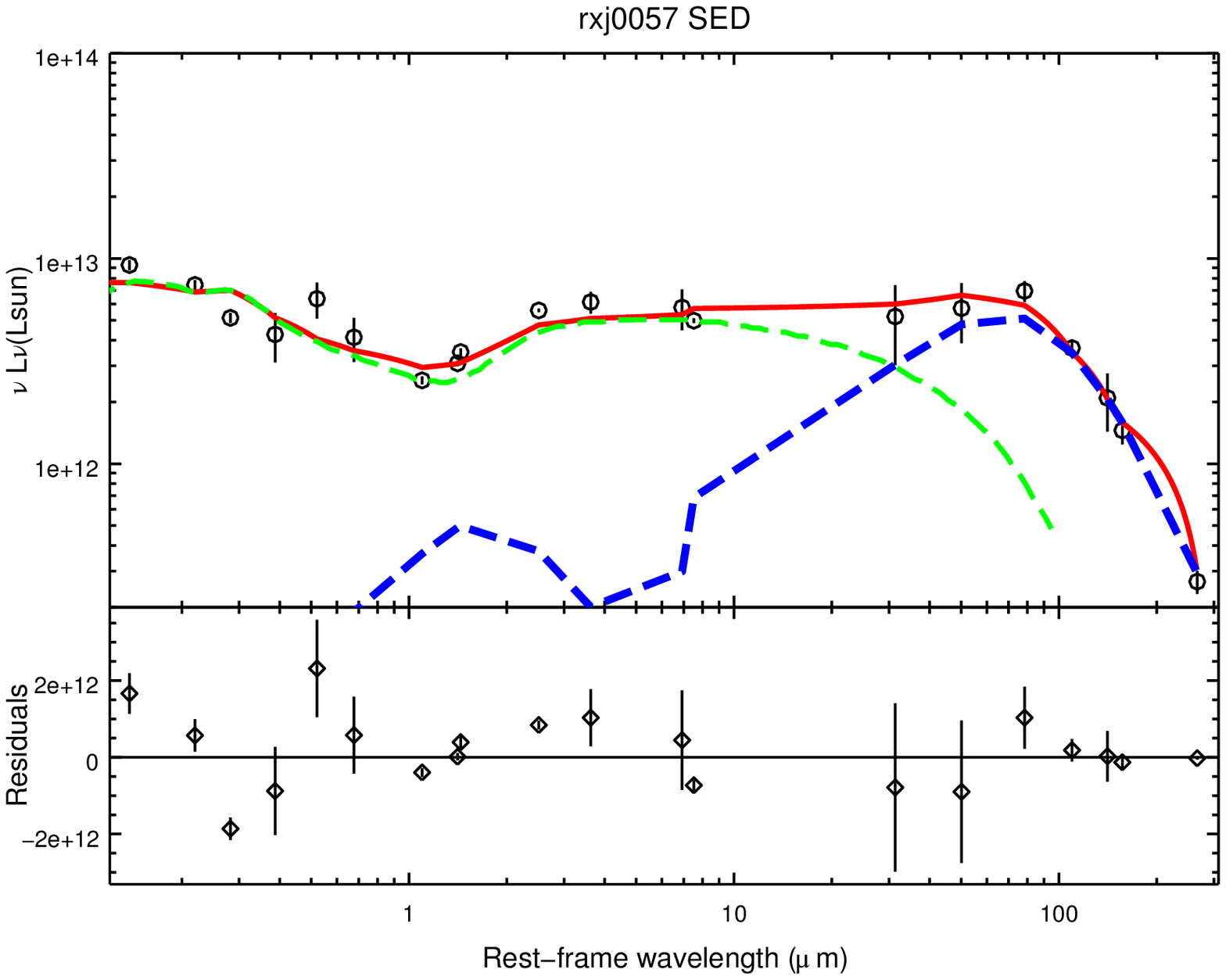}\\
   \includegraphics[width=0.49\textwidth]{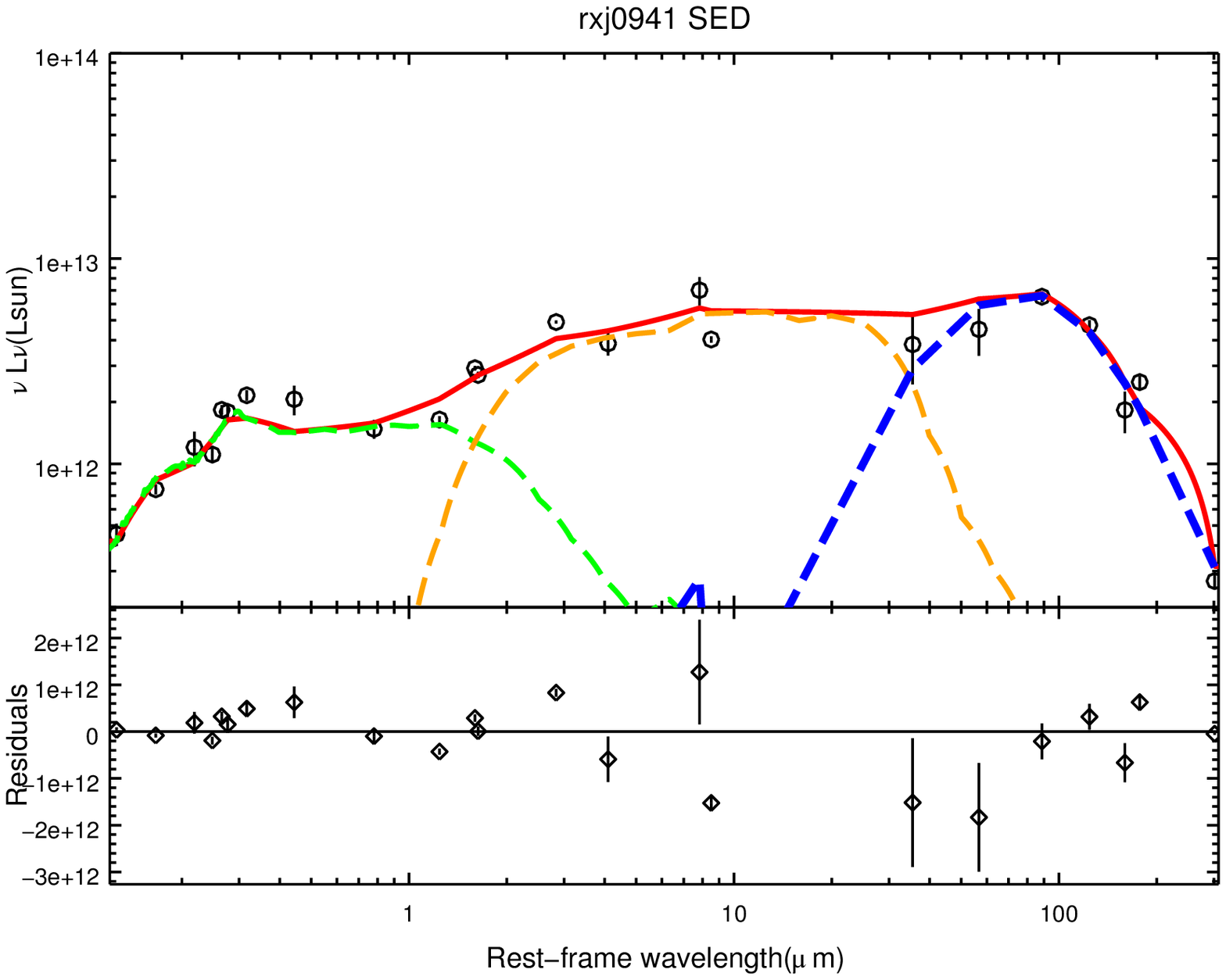}
   \includegraphics[width=0.49\textwidth]{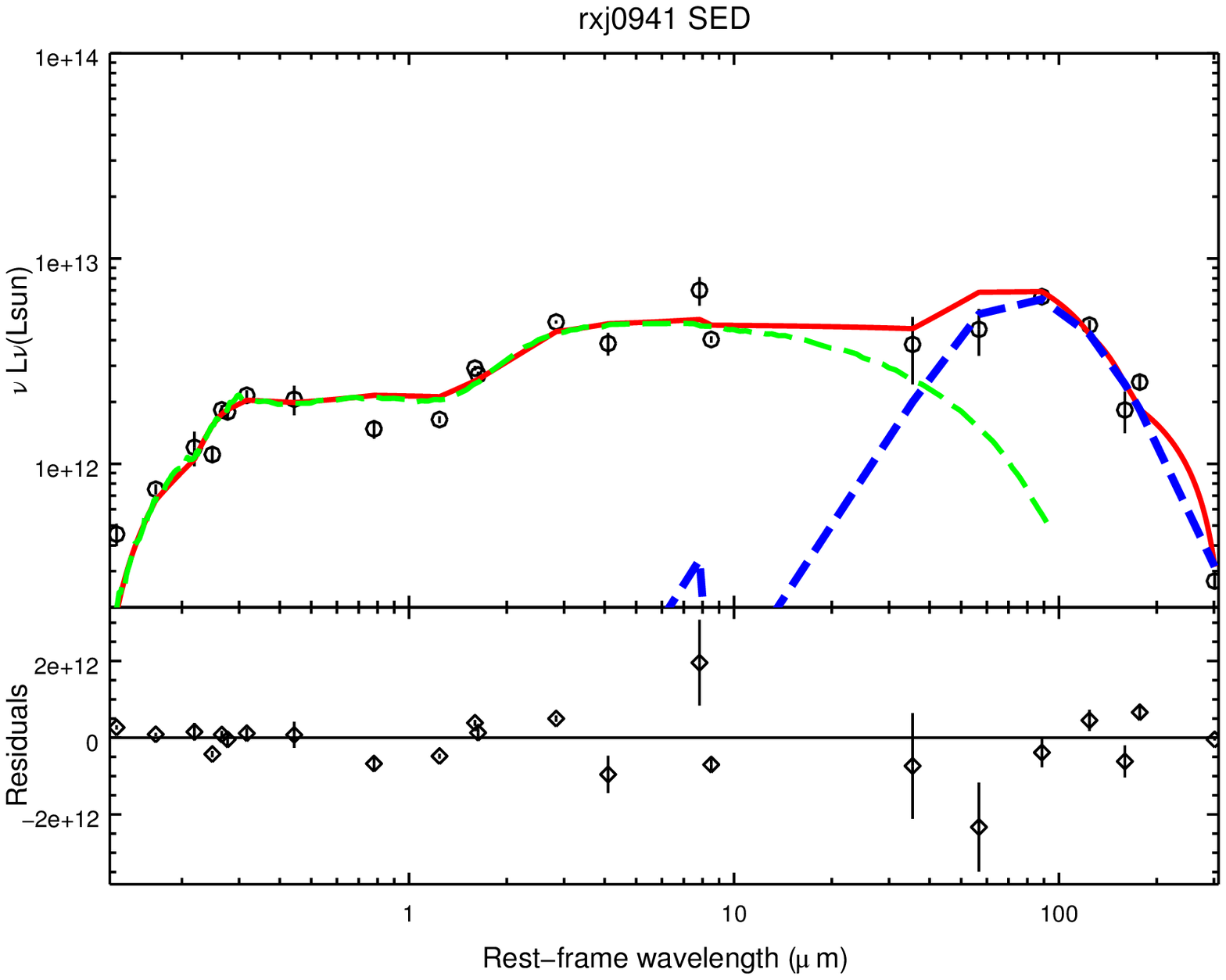}\\
   \includegraphics[width=0.49\textwidth]{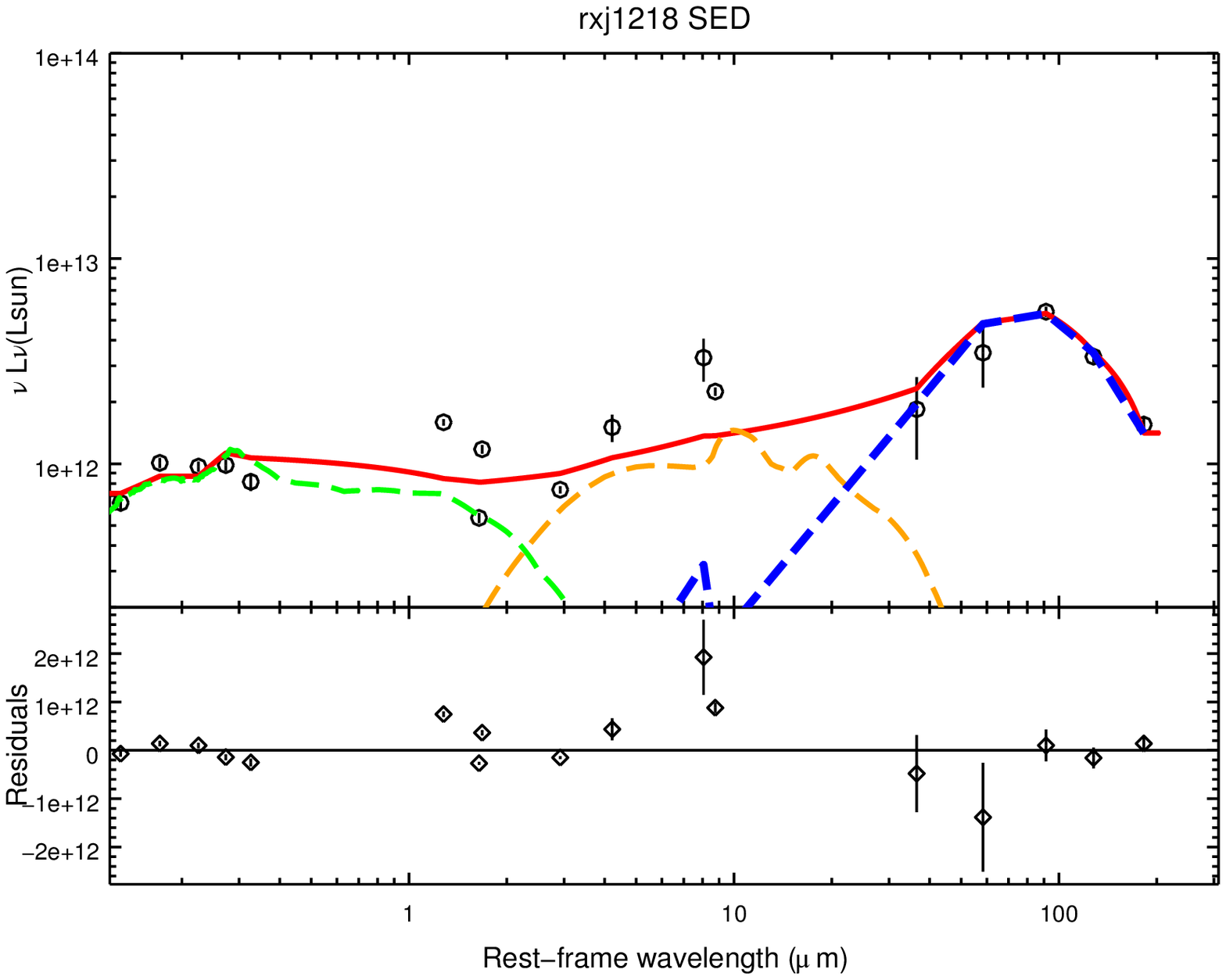}
   \includegraphics[width=0.49\textwidth]{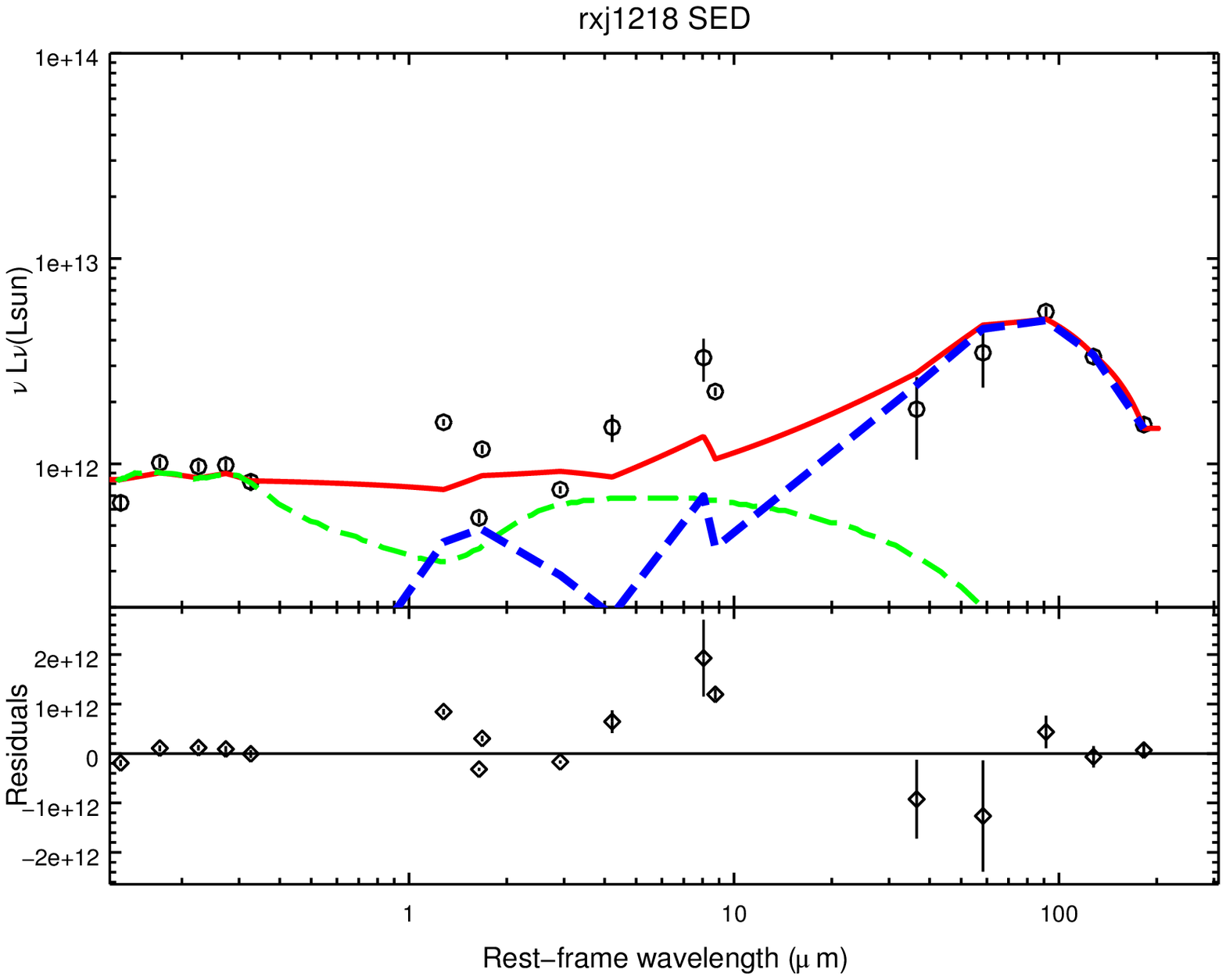}\\
   \caption{Best-fit SEDs to each source (one row per source) in
       two columns: on the left separate direct AGN (green dashed
       line), reprocessed AGN (torus, orange dashed line) and SF
       (blue dashed line) components; on the right, total AGN R06
       (green dashed line) and SF (blue dashed line) components. We
       also indicate for each source which is the best-fit torus model.
       First row: RX~J0057: Torus1. Second row: RX~J0941: Dusttor.
       Third row: RX~J1218: Torus2. Fourth row: RX~J1249: Torus1.
       Fifth row: RX~J1633: Torus3.}
  \label{figAllFits}
\end{figure*}

\begin{figure*}
   \includegraphics[width=0.49\textwidth]{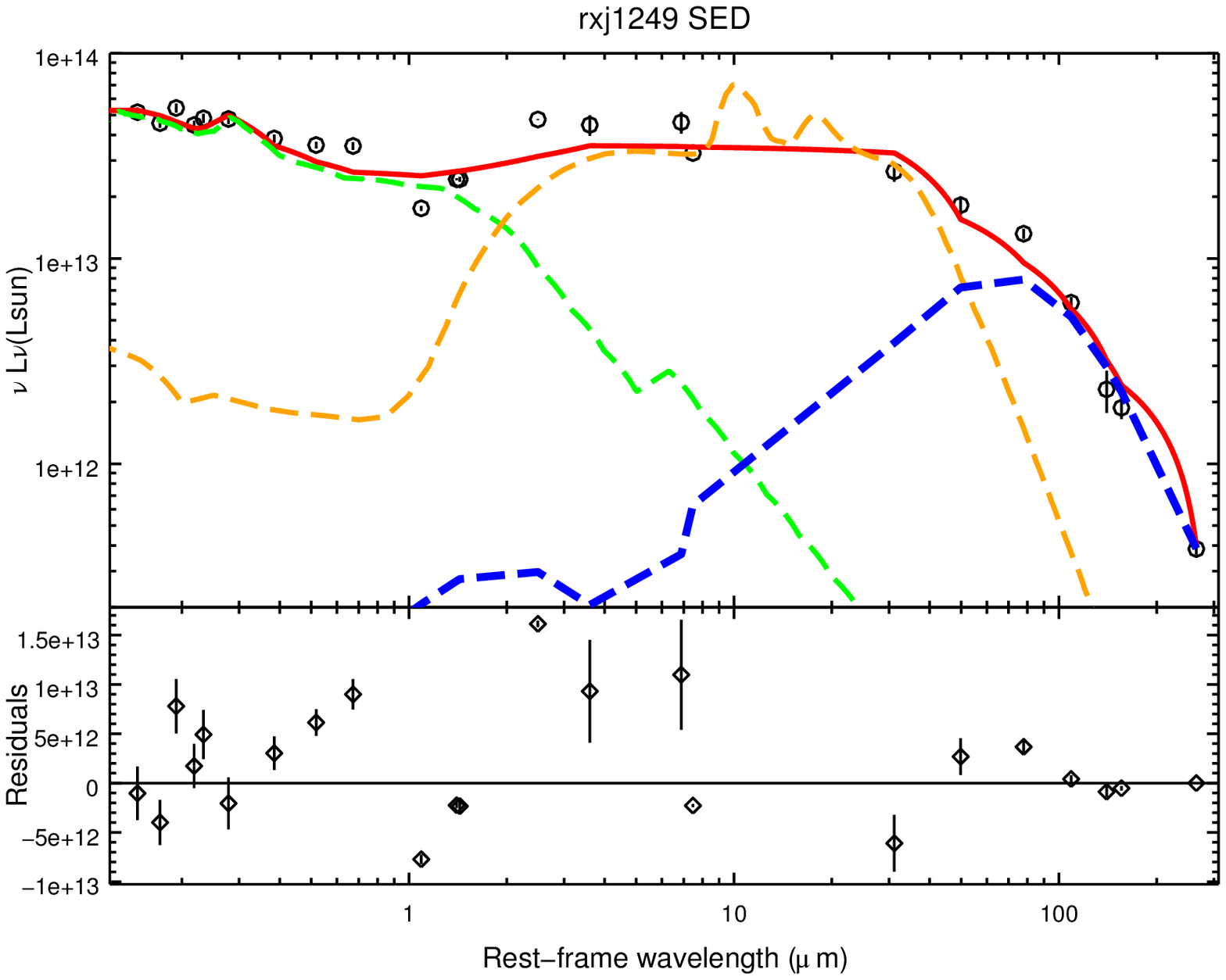}
   \includegraphics[width=0.49\textwidth]{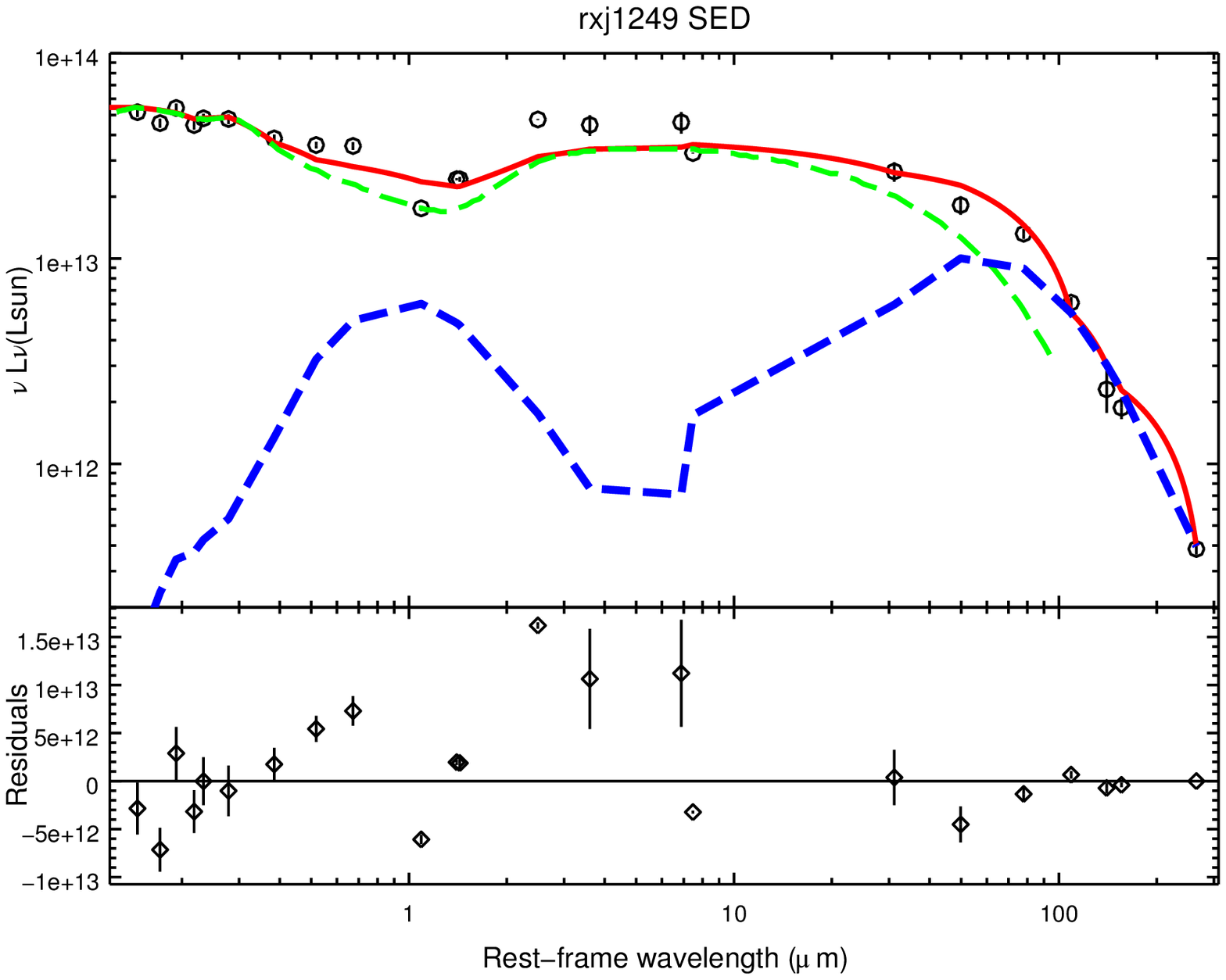}\\
   \includegraphics[width=0.49\textwidth]{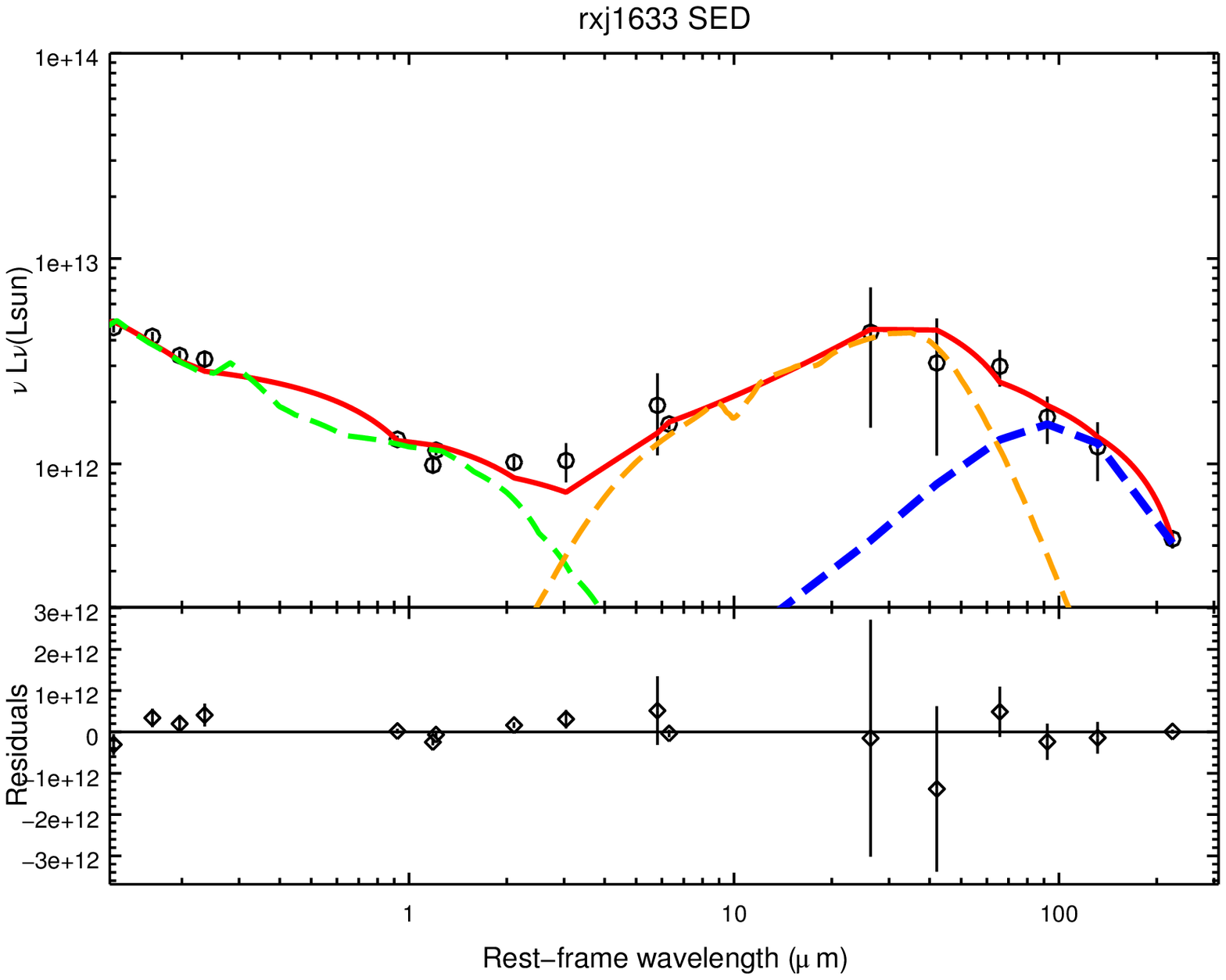}
   \includegraphics[width=0.49\textwidth]{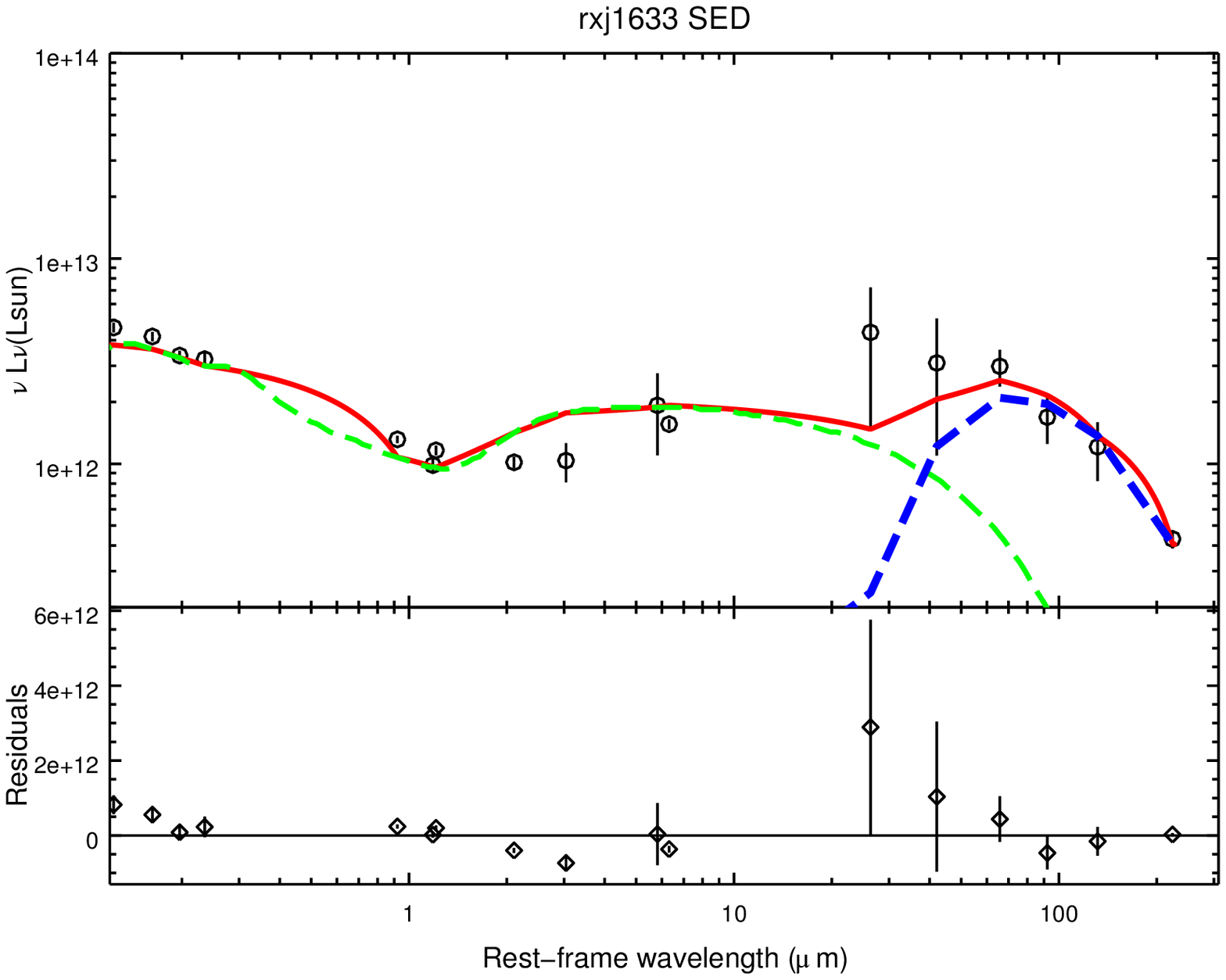}
   \contcaption{}
\end{figure*}

\begin{table*}
 \centering
 \caption{Average results and their uncertainties of fit parameters for each QSO, except for the X-ray AGN luminosity $L_{X,AGN}$ 
   which comes from \citealt{page11}. $\chi^2$ represents the minimum value for each best-fit family and $N$ represents the number of photometric points that we have for each object.}
 \begin{minipage}{180mm}
  \begin{tabular}{cccccccccc}
  \hline
  Object & $L_{X,AGN}$ &  $N_H$  & $L_{DISK}$ & $L_{TORUS}$  & $L_{FIR}$ & $L_{IR}$ & $T$ & $\beta$  & $\chi^2$ / N \\
  & ($10^{11} L_{\odot}$) &($10^{20}\, cm^{-2}$) & ($10^{11} L_{\odot}$) & ($10^{11} L_{\odot}$) & ($10^{11} L_{\odot}$) & ($10^{11} L_{\odot}$) & (K) & & \\
 \hline
 RX~J0057 & 2.93 & $0.01 \pm 0.05$ & $125 \pm 17 $& $190 \pm 30$  & $49 \pm 10$ & $170 \pm 40$ &  $38 \pm 4$ & $1.1 \pm 0.2$ & 159 / 20\\
 RX~J0941 & 0.93 & $0.77 \pm 0.11$ & $100 \pm 10$ & $150 \pm 20$ & $78 \pm 11$ & $180 \pm 30$  &  $36 \pm 3$ & $1.2 \pm 0.2$ & 347 / 23 \\
 RX~J1218 & 2.33 & $0.39 \pm 0.08$ & $42 \pm 7$ & $25 \pm 5$ & $63 \pm 6$  & $90 \pm 19$ & $36.0 \pm 1.7$& $1.2 \pm 0.2 $ & 414 / 17 \\
 RX~J1249 & 3.69 & $0.14 \pm 0.03$& $1230 \pm 90$ & $1160 \pm 100$& $70 \pm 15$ & $780 \pm 90$ &  $39 \pm 4$ & $1.2 \pm 0.2$ & 3109 / 23\\
 RX~J1633 & 5.85 & $0.003 \pm 0.002$ & $65 \pm 6$ & $70 \pm 30$  & $22\pm 5$ & $90 \pm 30$ &  $36 \pm 2$&  $1.00 \pm 0.11$ & 29 / 17\\
\hline
\end{tabular}
    \label{LumResults}
\end{minipage}
\begin{minipage}{180mm}
\end{minipage}
\end{table*}

\begin{table*}
 \centering
  \caption{Additional physical quantities derived from the average values and their uncertainties of the best-fit parameters.}
 \begin{minipage}{180mm}
  \begin{tabular}{ccccccccccc}
  \hline
  Object &  $L_{BOL}$ & CF &  $SFR$ &  $M_{DUST}$ &  $\Lambda_{EDD}$ & $\dot{M}$  & $k_{BOL}$\\ 
  & ($10^{11} L_{\odot}$) &  & ($M_{\odot}y^{-1}$)  & ($10^{9} M_{\odot}$) &   &\\ 
 \hline
 RX~J0057 & $250\pm 30$ & $0.78 \pm 0.15$ & $840 \pm 170$ & $2.5 \pm 0.8 $ & $0.09_{-0.07}^{+0.05}$ & $ 17\pm2$ & $85 \pm 10$\\ %& $ 48\pm 6 $
 RX~J0941 & $169 \pm 13$ & $0.87 \pm 0.14$ & $1350 \pm 190$ & $4.1 \pm 1.3$ & $0.09_{-0.08}^{+0.05}$ & $ 11.5\pm 0.9$ & $182 \pm 14$ \\ % & $ 118\pm 10 $
 RX~J1218 & $78 \pm 7$ & $0.32 \pm 0.07$ & $1090 \pm 100$ & $4.6\pm 1.6$ & $0.13_{-0.12}^{+0.08}$ &  $ 5.3\pm 0.5$ & $33 \pm 3$ \\ % & $ 210\pm 20 $ 
 RX~J1249 & $1890 \pm 90$ & $0.61 \pm 0.06$ & $1200 \pm 300$ & $3.8\pm 0.9$ & $0.61_{-0.6}^{+0.4}$ & $ 128\pm 6$ & $512 \pm 8$\\ % & $ 9.4\pm 1.6 $ 
 RX~J1633 & $186 \pm 15$ & $0.40 \pm 0.16$ & $380 \pm 90$ & $2.0 \pm 0.7$ &  $1.1_{-0.9}^{+0.6}$ & $ 12.6\pm1.0$ & $32 \pm 3$\\ %& $ 30\pm 3 $ 
\hline
\end{tabular}
    \label{discu}
\end{minipage}
\begin{minipage}{180mm}
\end{minipage}
\end{table*}

\subsection{Luminosities from SED fits}
\label{luminosities}

We show in Table~\ref{LumResults} the average value of the
8-1000~$\mu$m luminosity $L_{IR}$, calculated by integrating the torus
and SF templates in that range and adding both contributions.  
Its uncertainty has been calculated in a similar way to those of the
  best-fit luminosities (see above), using both the standard deviation
  among the best-fit families and the relevant photometric errors
  (using this time the 8-1000~$\mu$m range), adding them in
  quadrature.

We define $L_{BOL}$ as the integrated 100~keV-to-100~$\mu$m luminosity
from the AGN.  We estimate this quantity using the observed
absorption-corrected 2-10~keV (obtained from XMM-Newton spectra in
\citealt{page11}) and 0.12-100~$\mu$m luminosities of our objects. For
wavelengths longer than 0.12~$\mu$m we have integrated the fitted disk
template. At wavelengths shorter than 0.0035~$\mu$m (0.35~keV) we have
approximated the source continuum emission by a power-law
$L_\nu\propto\nu^{-1}$, normalized to the observed
absorption-corrected 2-10~keV luminosities \citep{page11}.  Between
those two wavelengths we have interpolated linearly in log-log
(equivalent to assuming a power-law shape). Most of the final
bolometric luminosity comes from the disk component (50-70 percent,
except for RX~J1633, for which this is 34 percent), so the exact
parametrization of the higher energies does not have a strong
influence on our results. We have calculated $L_{BOL}$ for each
best-fit family, showing in Table~\ref{discu} their average and 
its uncertainty, coming again from the dispersion among the best-fit
  families and the average relative photometric errors (using the same
  range as the one for the direct AGN component), adding them in
  quadrature.

\subsection{Black Hole masses and related quantities from optical spectral fits}
\label{BHmass}
We have estimated the BH masses of our objects from the MgII and CIV
emission lines from the optical-UV spectra in \cite{page11}. The
latter is usually preferred in the literature over the former because
it presents lower complexity and wider scatter. Unfortunately, MgII is
only covered by our spectra for RX~J0941; for this source we have
estimated the BH mass using both lines, to gauge the difference
between using them.  There are a number of different parametrizations
of the BH mass as a function of the width of those lines and the
luminosity in neighbouring wavelengths. We have chosen those used in
\cite{shen}:

\begin{eqnarray}
\label{eqMgII}
\log\left( {M_{BH} \over M_\odot} \right) = 2\log\left({FWHM \over 1000\,\mathrm{km/s}}\right)+ \nonumber \\
+0.50\log\left({\lambda L_\lambda\over 10^{44}\,\mathrm{erg/s}}\right)+6.86\,
\end{eqnarray}

\noindent for MgII \citep{vestergaad2009}, and

\begin{eqnarray}
\label{eqCIV}
\log\left( {M_{BH} \over M_\odot} \right) = 2\log\left({FWHM \over 1000\,\mathrm{km/s}}\right)+ \nonumber \\
+0.53\log\left({\lambda L_\lambda\over 10^{44}\,\mathrm{erg/s}}\right)+6.66\,
\end{eqnarray}

\noindent for CIV \citep{vestergaad2006}, with rms values of 0.55 and 0.36~dex, respectively.

\begin{table}
 \centering
 \caption{Black hole masses from fits to the CIV line at 1548\AA (except for
   the row marked with *, which comes from a fit to MgII at 2798\AA), FWHM
   corresponding with each line, $\lambda F_\lambda$ the monochromatic
   luminosity corresponding with each line and the Eddington luminosity.}
  \begin{tabular}{lcccc}
  \hline
  Object &  FWHM  & $\lambda F_\lambda$ & $\log(M_{BH})$ & $\log(L_{EDD})$ \\
         &  (\AA) & ($10^{44}$erg/s)             & ($\log(M_{\odot})$ & ($\log(L_\odot)$) \\
 \hline
RX~J0057  & 146$\pm$3     & 419$\pm$ 18  & 9.94$\pm$0.36 & 14.45$\pm$0.36 \\
RX~J0941  & 113$\pm$24    & 322 $\pm$29  & 9.77$\pm$0.40 & 14.27$\pm$0.40 \\
RX~J0941* & 106$\pm$7     & 202 $\pm$18  & 9.22$\pm$0.55 & 13.72$\pm$0.55 \\
RX~J1218  &  77$\pm$23    & 150 $\pm$28  & 9.28$\pm$0.45 & 13.79$\pm$0.45 \\
RX~J1249  &  86$\pm$26    & 3731$\pm$392 & 9.99$\pm$0.45 & 14.49$\pm$0.45 \\
RX~J1633  &  53.9$\pm$1.2 & 178 $\pm$6   & 8.73$\pm$0.36 & 13.24$\pm$0.36 \\
 \hline
\end{tabular}
\label{TabMBH}
\end{table}

\begin{figure}
   \centering
   \includegraphics[width=0.49\textwidth]{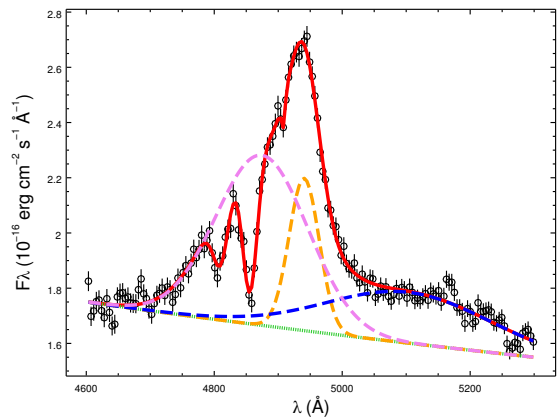}
   \caption{An example of the fits made to estimate the FWHM of the CIV emission line. 
   We have modelled this feature for RX~J0057 with three additive Gaussians 
   (blue, orange and violet dashed lines), a power-law continuum (dotted green line) 
   and three absorbing Gaussians (not shown individually). The total best-fit model 
   is shown as a red solid line.}
  \label{fig_fitFWHM}
\end{figure}

To obtain the $FWHM$ of the MgII and CIV lines, we have followed the
technique outlined in Shen et al. (2011), which we summarize here. For
MgII we have fitted, over the 2700-2900~\AA \ rest-frame range, one
narrow Gaussian component with $FWHM({\mathrm
  km/s})=\max(\Delta\lambda,500)$ where $\Delta\lambda$ is the
spectral resolution as given in \cite{page11} and one broad component,
plus four Fe ``humps'' at rest-frame wavelengths of 2630, 2740, 2886
and 2950~\AA. The continuum has been modelled as the best-fit
power-law over the rest-frame range 2200-2700~\AA \ (since the
additional 2900-2090~\AA \ range is not covered in our spectrum of
RX~J0941). For CIV we have fitted three Gaussians (only one was
necessary for RX~J0941) over the range corresponding to the expected
position of CIV at the redshift of the source $\pm 20000\,{\mathrm
  km/s}$, forcing their central wavelengths to be in the range
1500-1600~\AA. The continuum has been modelled as the best-fit
power-law over the rest-frame ranges 1445-1465 and 1700-1705~\AA,
see Figure~\ref{fig_fitFWHM}.

In addition, the spectra of our sources presented a number of narrow
and broad absorption lines, which we have modelled as multiplicative
absorbing Gaussians $M(\lambda)=
\left(1-A\exp\left(-{(\lambda-\lambda_0)^2 \over 2
      \sigma^2}\right)\right)$ with $0\leq A\leq 1$. We needed 3, 5,
7, 3 and 6 such components for RX~J0057, RX~J0941, RX~J1218, RX~J1249
and RX~J1633 (respectively) around CIV, and 2 for RX~J0941 around
MgII.

For RX~J0941 we have obtained the $1-\sigma$ uncertainty on the $FWHM$
directly from that of the $\sigma$ of the single broad line fitted to
the MgII and CIV lines, obtained in turn from the $\Delta\chi^2=1$
interval around the best fit value. Unfortunately, for the rest of the
sources this is not so straightforward, since the CIV emission is
modelled as the sum of three Gaussians, and there is no analytic
expression relating the overall $FWHM$ to the best fit parameters of
those three Gaussians (nine in total: three each of central
wavelength, width and normalization). We have therefore sampled the
parameter space, fixing those nine parameters to values around the
best-fit ones, calculating the best-fit to each spectrum in the above
range with the above constraints, calculating the $FWHM_i$ for each of
those combinations of parameters, and assigning to each a probability
$P_i=e^{(-\Delta\chi^2/2)}$ (Press et al. 1992). We have then
estimated the variance on the $FWHM$ as $\sigma^2_{FWHM}=\langle FWHM
-\langle FWHM \rangle \rangle \sim \sum_i P'_i(FWHM_i-\langle
FWHM\rangle)^2$ where $\langle FWHM\rangle\sim\sum_i P'_i FWHM_i$ and
$P'_i=P_i/\sum_i P_i$ (to normalize the total probability to unity).
For the sampling of each parameter we have used a Gaussian centred on
the best-fit value with a dispersion set to (in decreasing order of
choice) either the corresponding value of the covariance matrix (if
defined), or to the $\Delta\chi^2=1$ interval (if defined), or to 10\%
of the best-fit value (1\% in the case of the central wavelengths).
The latter percentages were the rough averages of the dispersions of
the similar parameters when they were defined.  In any case, we
checked that the final result did not depend on the particular choice
of these dispersions, since they only serve to increase the efficiency
of the sampling, favouring values close to the best-fit, where the
highest probabilities should concentrate. We sampled 2000 combinations
of parameters for each spectrum. We also checked that the dispersions
did not depend on the exact number of sampling points.

The estimates of the continuum values $\nu L_{\nu}$ in the Eqs. \ref{eqMgII}, \ref{eqCIV} (at rest-frame wavelengths of
1350 and 3000\AA \ for CIV and MgII, respectively) have been obtained
from the disk component that best fitted the overall SED of each
source (see Section \ref{fits}), which should provide a good estimate
of the intrinsic disk emission, corrected for absorption. We have
estimated the $1-\sigma$ uncertainties on those values from the
dispersion in the values of the normalization of that component
$L_{DISK}$ (see Section \ref{fits}).

Finally, the uncertainties in $\log(M_{BH}/M_\odot)$ have been
estimated from those of $FWHM$ and of the continuum using the standard
error propagation rules on the equations above (\citealt{bevington}).
We have added in quadrature the rms values in the equations above to
get our rather conservative estimates of the total uncertainties.  The
BH mass estimates and the total uncertainties are given in Table
\ref{TabMBH}. For RX~J0941 we get $\log(M_{BH}/M_\odot)=9.8\pm0.4$
from CIV and $\log(M_{BH}/M_\odot)=9.2\pm0.6$ from MgII, which are
within their mutual errors.

From the mass of the black hole we can calculate the Eddington
luminosity using \cite{eddington}:
\begin{equation}
\label{edd_eq}
L_{EDD} = 3.2\times 10^4\frac{M_{BH}}{M_{\odot}} L_{\odot}
\end{equation}

\subsection{AGN-related quantities from SED fits}
\label{diretAgenValues}
Once we have $L_{BOL}$ we can estimate naively the covering factor. Due to the anisotropy
of the torus radiation we can only estimate the bolometric covering factor 
or apparent covering factor \citep{mor2009,honig14}
$CF$ as the ratio between the reprocessed emission from the torus and
the total AGN bolometric luminosity:
\begin{equation}
\label{cf_eq}
CF = \frac{L_{Torus}}{L_{BOL}},
\end{equation}

This gives us a measure of how much of the sky seen from the central
source is intercepted by the torus. The values, shown in
Table~\ref{discu}, have been calculated from the average values
  of $L_{TORUS}$ and $L_{BOL}$ and their uncertainties, using the
  standard propagation of errors.

The Eddington ratio would then be $\Lambda_{EDD} = L_{BOL} / L_{EDD}$.
We show the values of $\Lambda_{EDD}$ in Table~\ref{discu}; they span
the range $\sim 0.1-1.1$. The error bars on $L_{BOL}$ are symmetric in
linear space, while those of $\log M_{BH}$ are large and symmetric in
log space.  The usual propagation of errors rules are only formally
valid for small errors and using them would result in symmetric error
bars for $\Lambda_{EDD}$ larger than the actual values. This does not
make sense, since both the bolometric luminosity and the black hole
mass are very significantly detected. To avoid these difficulties, we
have estimated the uncertainties on the Eddington ratios sampling
10000 times the distribution of $L_{BOL}$ and $\log M_{BH}$ using
Gaussians with the observed values, calculating the corresponding
Eddington ratios, and finding the narrowest interval that included
68.3 percent of the sampled values. The resulting asymmetric error
bars are those shown in Table~\ref{discu}.

\subsection{Star-formation-related quantities from SED fits}
\label{SFRsection}
Our SEDs require a contribution from cold dust, quantified by $L_{FIR}$. 
We have used the luminosities obtained in the previous Sections (given in tables \ref{luminosities} and \ref{discu}) 
to check whether that cold dust could be powered by the AGN, following a simple energetic argument. 
We have compared $L_{BOL}$ (the total AGN power available) with $L_{TORUS}$ 
(AGN power already intercepted by the ''warm'' reprocessing medium -the torus-) and $L_{FIR}$ 
(the observed power from cold dust): if $L_{BOL}$ is significantly larger than the sum 
of $L_{TORUS}$ and $L_{FIR}$, the AGN would have enough power to generate all the 
observed emission (with some to spare for the, mostly unobscured, observed UV-to-optical range). 
This can also be verified visually by looking at the SEDs in Fig.~\ref{figAllFits}. 
It is clear that the AGN emission would be insufficient in RX~J0057, RX~J0941 and RX~J1218. We have already 
discussed that the preferred torus model for RX~J1633 includes a larger contribution from colder gas, 
producing a lower LFIR, so some of the putative AGN contribution would be already corrected for. 
As we will see below, RX~J1249 is an exceptionally 
luminous QSO. However, its FIR emission is comparable to that of the other members of our sample, 
which would be unexpected it a significant contribution to that range came from such a powerful AGN. 
Concluding, the FIR emission observed in our objects likely predominantly comes from SF, 
as obviously it's not just star-formation.

The SFR has been calculated from $L_{FIR}$ using the following
expression from \cite{kennicutt},
\begin{equation}
\label{SFR_eq}
SFR(M_\odot/\mathrm{ y})=1.7217\times 10^{-10}L_\mathrm{ FIR}(L_\odot),
\end{equation}

We obtain values of $SFR\sim1000 M_\odot/\mathrm{ y}$ (shown in
Table~\ref{discu}). We confirm then our qualitative impression from
Fig,~\ref{figAllSEDs}: these objects are forming stars copiously.

Supernovae and binary stars associated with star formation produce
X-rays. \cite{myrto11} found a tight correlation between X-ray
luminosity and IR luminosity from ULIRGs, which can be expressed as
$L_{X,SF}=(1.9\times 10^{26}) L_{IR}^{0.3} + (4.15\times 10^{-5})
L_{IR}$ (their equation 3). For the typical values of our QSOs of
$L_{IR}\sim10^{46}$, this corresponds to $L_{X,SF} \sim
10^{42}$~erg/s. Comparing these values with the AGN X-ray luminosities
in the same band ($L_{X-AGN} \sim 4-20\times10^{44}$~erg/s), we conclude
that the contribution of the SF to the X-ray luminosity is very small
and can be neglected.

In order to estimate the values of the dust mass associated with star
formation, we have used the values of $L_{FIR}$, and the $T$ and
$\beta$ obtained in Section~\ref{fits} for each of the ``good''
sub-grids. We have related those three quantities to the dust mass
$M_{DUST}$ using Equations 2 and 5 in \cite{martinez}. Their Equation
2 can be written as
\begin{equation}
\label{eqLFIR}
L'_\mathrm{FIR}={ 8\pi h \over c^2} { d_L^2(z) F_{\nu'} \over (1+z) (\nu'(1+z))^\beta B_\nu(T,\nu'(1+z))}{1\over F}
\end{equation}
\noindent where $F_{\nu'}$ is the observed monochromatic flux at some
reference observed frequency $\nu'$ (in their case $\lambda'=1.2$~mm),
where
\begin{equation}
\label{eqF}
F=\int_0^\infty d\nu \frac{\nu^{3+ \beta}}{\exp(h \nu/kT)-1}=
{ \Gamma(\beta+4)\zeta(\beta+4)\over \left({h \over kT}\right)^{\beta+4} }
\end{equation}
\noindent (\citep{gradshteyn} FI II 792a) is the
normalization of the greybody function to the full frequency range.
Eq.~\ref{eqLFIR} is simply the total luminosity of a greybody
normalized to the observed value of $F_{\nu'}$. We have called this
quantity $L'_{FIR}$ to emphasize the difference between this and the
usual definition of $L_{FIR}$ over the 40 to 500~$\mu$m range.

Similarly, Eq. 5 in \cite{martinez} can be written as
\begin{equation}
\label{eqMdust}
M_\mathrm{ DUST}={ d_L^2(z) F_{\nu'} \over (1+z) \kappa(\nu'(1+z)) B_\nu(T,\nu'(1+z))},
\end{equation}
\noindent where 
\begin{equation}
\kappa(\nu(1+z))=\kappa_0\times\left({\nu(1+z) \over \nu'}\right)^\beta,
\end{equation}
We have used $\kappa_0=0.04$~m$^2$/kg at $\lambda'=$1.2~mm
\citep{beelen}. If we now divide Eq.~\ref{eqMdust} by Eq.~\ref{eqLFIR}, substituting Eq.~\ref{eqF}, we get
\begin{equation}
M_{DUST}={c^2 \over 8\pi h} {1\over \kappa_0\nu'^{-\beta}}{L'_{FIR}\over \int_0^\infty d\nu \frac{\nu^{3+ \beta}}{\exp(h \nu/kT)-1}}
\end{equation} 
\noindent where the last fraction is simply a normalization over the
full frequency range.

We have instead used
\begin{equation}
M_{DUST}={c^2 \over 8\pi h} {1\over \kappa_0\nu'^{-\beta}}{L_{FIR}\over \int_{40\,\mu\mathrm{m}}^{500\,\mu\mathrm{m}} d\nu \frac{\nu^{3+ \beta}}{\exp(h \nu/kT)-1}}
\end{equation} 
\noindent using the usual definition of $L_{FIR}$, which we have
obtained directly from our SED fits. We show in Table~\ref{discu} the
averages and standard deviations of the acceptable families of models.
Our approach involves two approximations: replacing the
renormalization over the full interval to that over the 40 to
500~$\mu$m interval, and replacing the FIR luminosity from a greybody
with that from a fit to a more sophisticated SF model. We believe that
we are justified to do both because, on the one hand, the greybody
shape decreases very fast outside the usual FIR range and, on the
other hand, the integrals over that range of the greybody fits to the
SF templates gave very similar values to those of the actual
templates.

We can also estimate the gas mass present in the starforming regions of
those galaxies from the dust masses, assuming a gas-to-dust ratio of
54, deduced by \cite{kovacs} for $z=1-3$ SMG (with an uncertainty of
about 20 per cent). This value is similar to the one obtained by
\cite{Seaquist} for the central regions of nearby submm bright galaxies. For
our typical dust mass of $\sim 10^9 M_{\odot}$ this corresponds to a
gas mass of $M_{GAS} \sim 10^{10}-10^{11}\, M_{\odot}$.

\subsection{Time scales}
\label{timescales}

Assuming an efficiency in conversion of accreted mass into radiation
$\eta$ (we have assumed 10\% e.g. \citealt{treister}) the mass accretion rate
$\dot{M}$ can be related to the bolometric luminosity $L_{BOL}$ by:
\begin{equation}
 \label{acr}
 L_{BOL} = \eta \dot{M} c^2
\end{equation}
\noindent The relative speed of galaxy-building through star formation
and black hole growth through accretion can be estimated from
$SFR/\dot{M}$. Both quantities are included in Table~\ref{discu}, with
errors derived using the standard propagation of errors.

Assuming a constant accretion rate, we can estimate the black hole mass
doubling time $\tau$ as
\begin{equation}
 \label{tdouble}
 \tau \sim \frac{M_{BH}}{\dot{M}}
\end{equation}
\noindent and, defining the maximum mass of the black hole as
$M_{BH,max}$ (which we take as $2 \times 10^{10}\,M_\odot$, the
maximum value of the observed black hole masses in the local Universe,
NGC 4889, from column 10 in Table 2 of \citealt{kormendy},
\citealt{mcConnell}), the time $\tau_{max}$ needed to reach it is
\begin{equation}
 \label{tmax}
 \tau_{max} \sim \frac{M_{BH, max}- M_{BH}}{\dot{M}}
\end{equation}
\noindent Alternatively, we could have used in the numerator of 
Eq.~\ref{tmax} simply $M_{BH, max}$, but the masses of the black holes in
the centres of our objects are already comparable to the maximum local
black hole mass, so Eq.~\ref{tmax} is more accurate. We have again
sampled the distribution of $\dot{M}$ and $\log M_{BH}$ to estimate
the uncertainties in these two timescales, as discussed above.

Assuming again a maximum local black hole mass, we can estimate what
would be its corresponding maximum host galaxy mass, $M_{BULGE,
  max}=8.53 \times 10^{12}\,M_\odot$, using the \cite{marconi}
relation. From that value and the current SFR observed in our objects,
assuming again constant SFR, we can estimate the time $\tau_{SB}$
needed to reach that maximum mass value as:
\begin{equation}
 \label{timeSB}
 \tau_{SB} \sim \frac{M_{BULGE, max}}{SFR}
\end{equation}
The values of all these timescales are shown in Table~\ref{times}, along with the time between their epochs and now (``look-back time''    \footnote{From \tt{http://www.astro.ucla.edu/wright/CosmoCalc.html}}).

\begin{table}
  \caption{Summary of timescales for QSOs and star formation, assuming
    constant mass accretion rates and SFR. $\tau$
    is the black hole mass-doubling timescale, $\tau_{max}$ is the time needed
    to reach the maximum local black hole mass (see text), and $\tau_{SB}$ is
    the time needed to reach the corresponding maximum host galaxy mass. $t$ is the ``look-back time'' (see text).}

  \begin{tabular}{ccccc}
  \hline
  Object &  $\tau$ & $\tau_{max}$ & $\tau_{SB}$ & $t$\\
        &  $(Gy)$ & $(Gy)$ & $(Gy)$ & $(Gy)$ \\
 \hline
 RX~J0057 & $0.5_{-0.4}^{+0.3}$ & $0.7_{-0.4}^{+0.4}$ & $10 \pm 2$ & 10.6 \\
 RX~J0941 & $0.5_{-0.5}^{+0.3}$ & $1.2_{-0.4}^{+0.5}$ & $6.3 \pm 0.9$ & 10.0 \\
 RX~J1218 & $0.4_{-0.3}^{+0.2}$ & $3.4_{-0.5}^{+0.5}$ & $7.8 \pm 0.7$ &  9.9 \\
 RX~J1249 & $0.08_{-0.07}^{+0.05}$ & $0.08_{-0.06}^{+0.07}$ & $7.1 \pm 1.8$ & 10.6 \\
 RX~J1633 & $0.04_{-0.04}^{+0.02}$ & $1.55_{-0.15}^{+0.11}$ & $22 \pm 5$ & 11.3 \\
\hline
\end{tabular}
  \label{times}
\end{table}

\section{Discussion}

In this Section we will piece together the clues obtained above about
the nature of our objects.

From the SED shapes and template fits, we see some differences in the properties of the objects in our sample: 
RX~J0057 and RX~J1249 have similar redshifts
$z\sim2.2$, small optical obscurations, the highest bolometric AGN
luminosities and a preference for the first torus model in
\cite{roseboom}. RX~J0941 and RX~J1218 have again similar redshifts
$z\sim1.8$, higher optical obscurations, the lowest bolometric AGN
luminosities and also allow for the empirical \cite{rowan} and the
second \cite{roseboom} torus models.  Finally, RX~J1633 has the
highest redshift $z=2.8$, small optical obscuration, intermediate
bolometric AGN luminosity and a strong preference for the third torus
model in \cite{roseboom}. This results in a higher torus contribution at longer
wavelengths, intuitively corresponding to a higher probability of
further reprocessing of the direct emission at larger distances and
lower temperatures. Additional evidence for a (relative) higher
inclination of RX~J1633 comes from its radio morphology in FIRST
images, which shows two diffuse blobs roughly symmetrical with respect
to the optical position of the QSO, in contrast with the rest of our
objects, which are either not detected (RX~J0057 and RX~J1218) or
pointlike (RX~J0941 and RX~J1249).

The host galaxies of RX~J0941 and RX~J1218 seem to be growing much faster than their black holes, both
in absolute terms and compared to the other objects. This is because
their AGN are less luminous but have similar $SFR$ to the rest. The
difference in AGN luminosities between the objects could
be ascribed to the redshift, since those further away are also more
luminous, as expected in a flux-limited sample, but this is belied by
the intermediate luminosity of the highest redshift object RX~J1633,
which is also the one with the most discordant torus properties. In
any case, with such a small sample it is impossible to assess the
significance and implications of the differences observed.

\subsection{AGN properties}
In this section we want to study the main properties of the AGN.
Taking the sample as a whole, we can compare their overall properties
to those of type 1 QSO from the SDSS DR7 \citep{sdss7}: bolometric
luminosities \citep{shen}, Eddington luminosities and ratios
\citep{kelly} and covering factors \citep{roseboom}.

As expected from the design of the sample (\citealt{page2000}), their
X-ray and bolometric luminosities are similar to those of QSOs in
their redshift ranges (using a range $\Delta z=0.4$), except for
RX~J1249, which is truly exceptional: there are only 50 objects
brighter than RX~J1249 ($<$0.05\%) in the whole sample of \cite{shen}, 
and it is one of the most luminous object even comparing with the
  high-bolometric-luminosity-selected sample of \cite{tsai2014}.
Also, as discussed above, the shape of their SEDs in the optical-MIR
region is unexceptional.

From Table~\ref{discu}, three of our QSOs have Eddington ratios around
0.1 (RX~J0057, RX~J0941 and RX~J1218), one has a value around 0.6
(RX~J1249) and the last one is closer to unity (RX~J1633), although
with large error bars, specially in the last two. This range of values
is compatible with that found by \cite{kelly}, who have studied a
sample of $\sim$ 58000 Type 1 QSO, also finding that $\Lambda_{EDD}
\lesssim 0.1$ are significantly rarer at $z\sim 4$ compared to
$z\lesssim 2$, while objects with $\Lambda_{EDD} \gtrsim 0.1$ are
found similarly in both redshift ranges.  This is also similar to what
we see in our sample, where the four objects with $z\sim 2$ are
compatible with $\Lambda_{EDD} \sim 0.1$, while RX~J1633 at $z=2.8$
shows some preference for higher values of $\Lambda_{EDD}$.  They
conclude that Type 1 quasars radiating near the Eddington limited are
extremely rare, suggesting that Type 1 quasars violating the Eddington
limit do so only for a very brief period of time. RX~J1633 could be in
this situation.

\cite{roseboom} have estimated the covering factors $CF$ of their
WISE-UKIDSS-SDSS (WUS) quasar sample (5281 quasars) with $z<1.5$. They
have found an average value of $CF=0.39$, but in addition they also
get that two-thirds of type 1 quasars have $CF$ in the range 0.25 to
0.61, roughly as in our sample. Segregating the sources according to
their bolometric luminosities, our four lower luminosity objects are
in their $ 46 < \log (L_{BOL}) < 47$ bin. A quick MonteCarlo sampling
of the log-normal fit to their $CF$ distribution (calculating values
from our sample with respect to the mean values of the distribution
and five random distribution values with respect to the mean values
too) shows than an unremarkable 40~per cent of the simulated samples
have a similar distribution as our objects in this bin. In contrast,
RX~J1249 is in their highest luminosity bin ($\log (L_{BOL}) \sim
48$), where only three percent of the objects have a higher $CF$ than
this source: RX~J1249 is not only one of the most luminous objects
known, it also has an exceptionally high reprocessed-to-bolometric AGN
luminosity ratio for its luminosity.

We have calculated the bolometric corrections to the X-ray
luminosities for our sample defined as:
\begin{equation}
\label{kbol}
 k_{BOL} = \frac{L_{BOL}}{L_{X,2-10}}
\end{equation}
\noindent finding values $k_{BOL} \sim 30-500$ (see Table~\ref{discu})
for $\Lambda_{EDD}\sim 0.1-1$. Although at face value this looks very
different from the results of \cite{vasudevan}, our three objects with
$\Lambda_{EDD}\sim 0.1$ (RX~J0057, RX~J0941 and RX~J1218) are well
within the values spanned by the objects in their sample at similar
Eddington ratios $k_{BOL}\sim 30-180$ (see their Fig.~6). The most
discrepant objects are RX~J1249 and RX~J1633. The former presents an
extremely high bolometric correction around 500, more in line with its
BAL nature \citep{grupe08, morabito}. On the contrary, RX~J1633, our
highest redshift object, presents a bolometric correction of about 30
for an Eddington ratio of about 1. Only one object from
\cite{vasudevan} has a smaller correction for that ratio, although
there are a few more with $\Lambda_{EDD}>0.1$ and $k_{BOL}\lesssim 40$
in their sample, so it is not exceptional. We have also compared our
bolometric corrections with those of \cite{marconi04}. Only RX~J0057
is close to their relation and inside the dispersion region, RX~J1249
and RX~J0941 are above the relation and RX~J1218 and RX~J1633 are
below the curve.  Again the most discrepant sources are RX~J1249 and
RX~J1633.

\subsection{Star formation properties}

Regarding now the star formation properties of our sample, we have
obtained very high SFRs$\sim 1000 M_{\odot}$/y: these objects are
forming stars copiously and they are at the ULIRG/HLIRG level with
strong FIR emission.

The distribution of greybody temperatures ($T=36-39$~K) of the SF
templates of our sources seems to be closer to those of SMGs ($T \sim
35$~K, \citealt{beelen}), than to those of other high redshift QSOs
($T \sim47$~K, \citealt{kovacs}): the range of temperatures in the
latter work is $T\sim 40-60$~K while only two of our sources have
temperatures compatible with the lower bound of the interval (RX~J0057
and RX~J1249, the first subset discussed at the beginning of this
section). On the other hand, we obtain greybody slopes ($\beta=1-1.2$)
below those of typical SMG ($\beta=1.5$ \citealt{beelen}) and high
redshift QSOs ($\beta=1.6$ \citealt{kovacs}), even taking into account
the substantial error bars. However, our temperatures and slopes could
be affected by the limited wavelength range of the greybody fits to
the SF templates, since the SF thermal bump in our sources is
generally broader than in simple greybody models, and hence the fit
tends towards low values of $T$ (to match the steep decline at the
longest wavelengths) and low values of $\beta$, to accommodate the
extra width at the shorter wavelengths. If we just use a simple
greybody model to parametrize the SF, instead of the more elaborate
templates of \cite{siebenmorgenSB}, we confirm the above tendency:
both RX~J0057 and RX~J1249 prefer higher temperatures than in the
limited-range-fit to the best-fit SF template, while the rest show
similar or slightly lower temperatures. This would go in line with the
fact that in the former two objects the maximum of the SF emission is
lower than that of the torus emission, while in RX~J0941 and RX~J1218
the opposite happens.  Finally, it might well happen that a single
temperature fit without some modification at the high frequency end is
not sufficiently accurate to allow a precise determination of
$\beta$-T values. In any case, the exact values of those parameters in
particular fits have a very limited impact in our inferred
  luminosities, since we derive quantities from averages and
dispersions over the best group of fits.

\subsection{AGN-SF relationship}
\label{AGN-SF}

\begin{figure}
   \centering
   \includegraphics[width=0.52\textwidth]{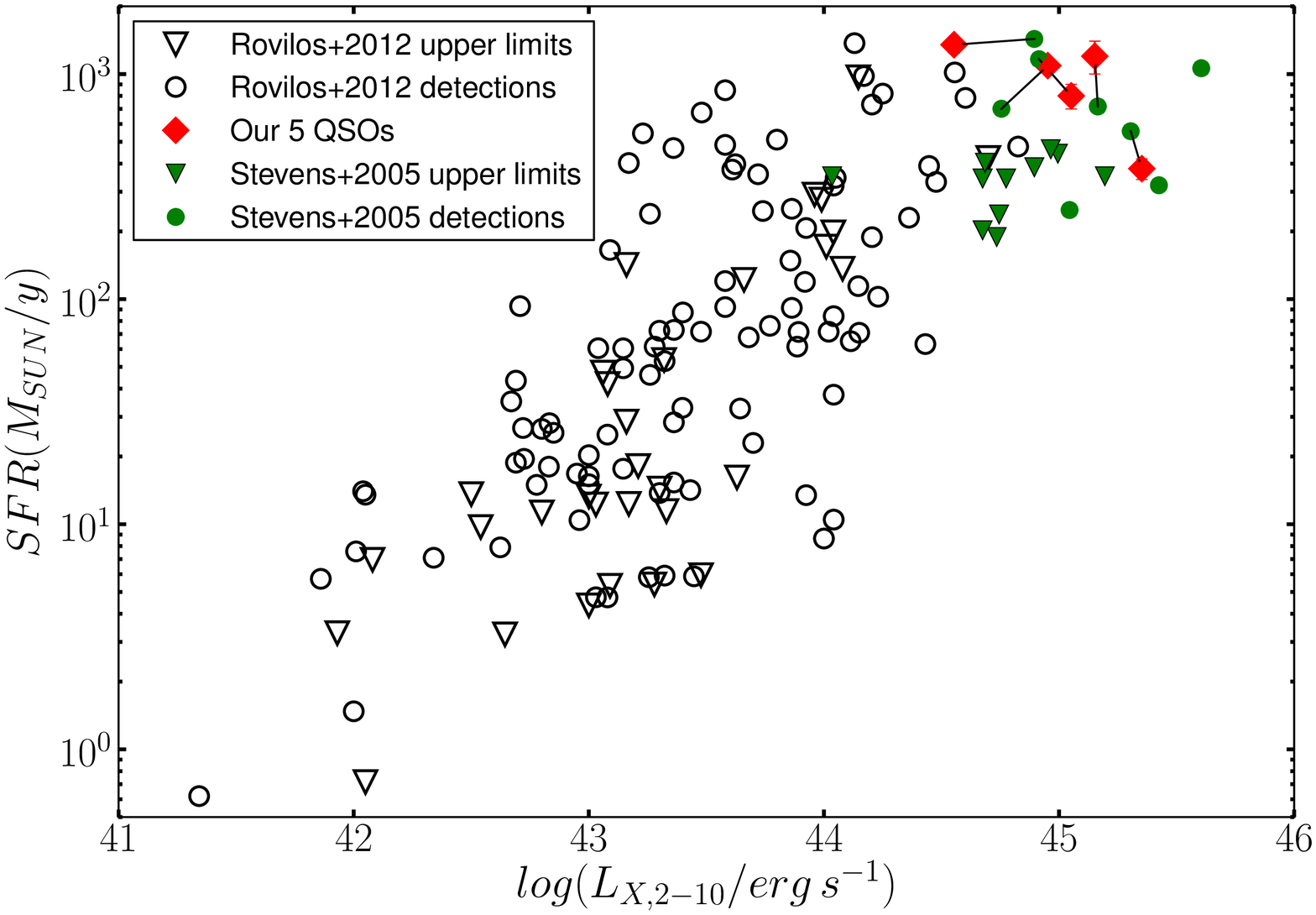}
   \caption{Star-formation rate versus X-ray luminosity for our five
     quasars (red solid diamonds, with values as reported in this
       paper).  For comparison, we also show X-ray selected AGN in
     deep Herschel surveys from \citealt{rovilos}: 99 FIR-detected
     (empty dots) and 32 with only FIR upper limits (empty triangles). 
     We have also included data from our full parent sample in
       \citealt{stevens05}, modifying the X-ray luminosities and SFR
       to a frame coherent with the data in this paper (see text):
       detected sources (green solid circles, joining with a segment
       the points corresponding to the common sources) and upper
       limits (green solid triangles).}
   \label{fig_rovilos}
\end{figure}

In the context of possible evidence for an influence of the central
AGN on the evolution of its host galaxy, it is interesting to compare
the rate of black hole growth (as gauged from e.g. the X-ray
luminosity $L_{X,2-10}$, the intrinsic bolometric luminosity from AGN
$L_{BOL}$, or the accretion rate $\dot{M}$) to the rate of galaxy
growth (from the SFR, the LIR or the monochromatic 60$\mu$m
luminosity).  \cite{lutz} studied the correlation between $log(\nu
L_{\nu}$) at 60$\mu$m and PAH emission (as proxies of the SF
luminosity) and $log(\nu L_{\nu}$) at 5100\AA (as a proxy for the AGN
luminosity) for a sample of QSOs at similar $z$ to ours with Spitzer
spectroscopic data in the rest-frame mid-infrared. They have found
that, at high luminosities and $z$, there was a flattening of the
relation between SF and AGN luminosity that is observed for lower
redshift QSOs. \cite{munalley2012} used 100 and 160 $\mu$m fluxes from
GOODS-Herschel finding no evidence of any correlation between the
X-ray and infrared luminosities of moderate luminosity AGNs at any
redshift. \cite{rovilos} found a significant ($>$99\%) correlation
between $L_{X,2-10}$ and SFR (see Figure~\ref{fig_rovilos}) in a deep
GOODS-XMM-Newton-Herschel sample, taking into account the upper limits
on the latter using the ASURV package (\citealt{asurv}). They also
found a significant correlation between the specific SFR (sSFR) and
X-ray luminosity, taking into account both upper limits and a possible
partial correlation with the redshift. \cite{rosario} used the
COSMOS-GOODS- North and South X-ray selected sample. They study
$log(\nu L_{\nu}$) at 60$\mu$m vs. $L_{BOL}$ finding a significant
correlation between $L_{\nu}$ at 60$\mu$m and $L_{BOL}$ for moderate
redshifts ($z<1$) and high luminosities.

Interestingly, all our objects are in the redshift range ($1<z<3$)
studied by \cite{page13}. They studied Herschel SPIRE observations of
the CDF-N field (within the HerMES project) and found evidence for
star formation (from 250~$\mu$m detections) in some X-ray detected AGN
with $10^{43}<L_{X,2-10}<10^{44}$~erg/s, but lower star formation in
the $10^{44}<L_{X,2-10}<10^{45}$~erg/s luminosity bin. They took this
as evidence for suppression of SFR in the most luminous AGN at that
epoch in the Universe. Since all our objects are very significantly detected
in 250~$\mu$m and have $L_{X,2-10}>10^{44}$~erg/s (see
Fig. \ref{fig_rovilos}), in this scenario these highly luminous QSOs 
would not yet have managed to switch off SF.

In this controversial context, we have revisited the $\log(SFR)$ vs.
  $\log(L_{X,2-10})$ correlation using both their sample and a joint
  sample with our sources (including data from \citep{stevens05}). We
  have tested for a ``hidden'' correlation with redshift (specifically
  with the luminosity distance $\log(d_L)$) using the method in
  \cite{akritas96}, who give their significance in terms of a
  ratio between the generalized Kendall's $\tau$ and its dispersion
  $\sigma$.

  We first started with just their sample,
  finding $\tau/ \sigma=5.1$. We
  tested this significance against simulated samples of sources with
  mutually uncorrelated SFR and X-ray luminosity values, but both
  correlated with redshift. Briefly, we found the constants $K$ that
  best reproduced the $\log(Y)=K+2\log(d_L)$ relations, with
  $Y=\log(SFR)$ (using only detections) and $Y=\log(L_{X,2-10})$,
  estimated the rms around these relations for several ranges of
  redshift, and then created 10000 samples of sources keeping the
  redshifts of the observed sources and simulating the X-ray
  luminosities and the SFR with the above ``calibrations''. For upper
  limits in SFR we kept the observed upper limit but randomized the
  X-ray luminosity as above. These simulated samples keep the
  statistics of the \cite{rovilos} sample but do not have any real
  correlation between SFR and X-ray luminosity. We found that 653 of
  those had $\tau/\sigma>5.1$, so we
  conclude that the real significance of the correlation in their
  sample is about $\sim 1-653/10000=93$\%$<2\sigma$.

  We now wish to add our sample to this relation to check its
  influence. We need to use our full parent sample in
  \cite{stevens05}, including the non-detections at 850~$\mu$m. 
  Note that there are three more ``detections'' in that paper
    compared to this paper: RX~J1107+72 (because it is radio-loud and
    hence its FIR emission originates in the AGN), RX~J0943+16 and
    RX~J1104+35 (these two are $<3\sigma$ significant). The 0.5-2~keV
  luminosities in that paper \citep{stevens05} have been converted to
  2-10~keV luminosities assuming $L_\nu\propto \nu^{-1}$ (we have
  checked that this method agreed well with the 2-10~keV luminosities
  from \citep{page11} used elsewhere in this work). For the five
  common sources, we have fitted the best multiplicative constant
  between the 850~$\mu$m-derived FIR luminosities in 
  \cite{stevens05} and our SED-derived FIR luminosities, taking
  into account the uncertainties in both quantities. Using this
    multiplicative constant we have then derived ``corrected'' FIR
    luminosities and SFR (EQ.~\ref{SFR_eq}) for all sources in
    \cite{stevens05} sample (green points in Fig.\ref{fig_rovilos}).
    The linked points show the magnitude of the re-scaling for the
    five common sources (red -this paper-, green -rescaled-). The
    rescaled full parent sample has been joined to the \cite{rovilos}
    sample to study the X-ray luminosity-SFR correlation.  Again, we
  have created 10000 random samples as explained above (re-calibrating
  the SFR and X-ray luminosity correlations with redshift and their
  rms).  The joint sample gave $\tau/\sigma=5.61$.  Only 23 out of
  10000 random simulations showed higher values, so the significance
  is $\sim$99.8\%$>3\sigma$. This joint correlation is more
  significant probably because our five detected sources extend the
  X-ray luminosity range by almost an order of magnitude.

  At face value, it appears that there is a significant correlation
  between the growth of galaxies via star formation and the growth of
  their central SMBH via accretion in the joint sample. However, a
  number of caveats are in order to interpret the observed
  correlation. First, given the very different selection functions of
  the samples involved (between them and among their constituent
  surveys at different wavelengths) and the small numbers of sources
  involved, it is very difficult to assess the significance of the
  different results in terms of the full population of
  black-hole-growing and/or star-forming galaxies at that epoch in the
  Universe. Large samples of objects at the relevant redshifts with
  well-controlled selection functions are needed to appraise this
  crucial issue. Furthermore, our SED-derived SFR cover very well the
  FIR rest-frame range even at our highest redshift (as they include
  observed frame points between 100 and 850~$\mu$m) and the
  ``corrected'' ones used for the correlation come originally from
  observed-frame 850~$\mu$m observations (similar to \citealt{lutz}).
  In contrast, the highest observed wavelength in \cite{rovilos},
  \cite{munalley2012} and \cite{rosario} is 160~$\mu$m, which only
  covers %at most the rising part
  the short-wavelength side of the FIR bump at redshifts above
  1, with the ensuing uncertainty in the FIR luminosities, despite the
  careful SED fit.

\begin{figure}
   \centering
   \includegraphics[width=0.49\textwidth]{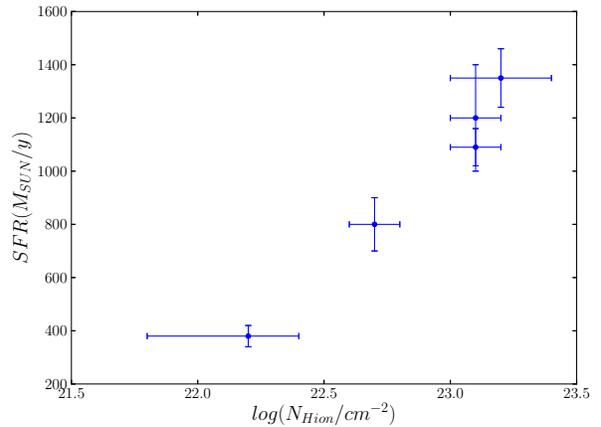}
   \caption{Star formation rate versus ionised column density
     N$_{H_{ion}}$ for our sample.  (from \citealt{page11}).}
   \label{fig_NH_SFR}
\end{figure}

Previous works have found some evidence of a correlation between the
SF of the host and the AGN obscuration by neutral gas in the
X-rays (\citealt{alexander}, \citealt{bauer}, \citealt{georgakakis},
\citealt{rovilos07}). Later studies with deeper surveys both in the
X-ray and infrared ranges have failed to reproduce these results (e.g
\citealt{rosario}, \citealt{rovilos}).  We have therefore looked 
instead for a correlation between the ionised column density 
N$_{H_{ion}}$ (from \citealt{page11}) and SFR in our sources (Fig.
\ref{fig_NH_SFR}), finding a tentative positive correlation between
these parameters.  This is interesting, since it would imply a
coupling of the ionized gas absorbing the X-rays at the scale of the
accretion disk or the BLR with the gas forming stars in the host
galaxy bulge, about three orders of magnitude farther away.  At face
value, this would be compatible with a positive feedback scenario,
\citep{king13} in which the ionized outflowing gas would trigger star
formation in the interstellar medium of the host galaxy, with the
highest column density gas corresponding to stronger feedback.

Alternatively, the AGN may also be ionising gas at kpc scales,
  co-located with the SF gas, so the above correlation would just be a
  consequence of the gas of higher density forming stars more
  intensely. Testing this intriguing possibility (with implications on
  positive and negative feedback) would need larger samples with a
  better determination of the location of the ionised gas.

\subsection{Evolutionary status}

We now turn to the fate of the objects in our sample. Assuming
constant accretion rates, we have estimated in
Section~\ref{timescales} their BH mass-doubling times $\tau$ and the
time they would take to reach the maximum BH mass observed locally
$\tau_{max}$. We note that RX~J0057, RX~J0941 and RX~J1249 are already
within a factor of 2-3 of this maximum local BH mass ($\sim 2\times
10^{10}\,M_\odot$), so their BH growth phase is expected to finish
within one to three mass-doubling times at most, i.e. 700, 1200 and 80
million years respectively. Given the expected lifetime of an active
QSO phase of about 200 million years \citep{hopkins}, RX~J1249 would
have time to reach that maximum mass, while the other two objects
would stop at a lower, but not much lower, mass. The BH in the centre
of RX~J1633, the least massive in our sample (but already with a
considerable mass of about $10^9\,M_\odot$, larger than about 60\% of
the objects in \citealt{marconi}), is growing quite fast, doubling its
mass in 40~million years, so it could increase its mass by a factor of
a few within the fiducial QSO lifetime above. RX~J1218 has a
intermediate BH mass in our sample (about 10 times lower than
the local maximum), but its accretion rate is the lowest in the
sample, so its mass-doubling time is 3~Gy. Therefore, unless something
happens to re-kindle the AGN at a later time, it is unlikely to grow
much more.

Without a determination of the masses of the host galaxies of our
objects we cannot perform a similar exercise for the expected lifetime
of the starburst phase. Unfortunately, the UV-to-MIR range is
completely dominated by the AGN light. We are trying to secure mm
observations to estimate the host galaxy masses directly.
Nevertheless, we can argue that the SFRs observed are very high
(albeit with considerable uncertainties), among the highest observed
at the relevant redshifts, so they are unlikely to be maintained for a
long time.

Considering further episodes of strong galaxy growth with little AGN
growth does not help to escape that the galaxies should be already
formed, since either the host galaxy has to swallow whole fully-formed
large-mass galaxies without forming stars (and with insignificant BH
masses), or the ensuing SFR would have to be several times larger than
the already extraordinary ones found in our objects (while at the same
time avoiding a significant infall of gas to the galaxy nucleus to
make sure to starve the massive BH already in place).

In an independent line of argument, in the event that there is a
  positive feedback at work, as discussed in Section \ref{AGN-SF},
  then the bulk of the gas mass in the host galaxies could be about to
  form stars pressured by the now maximal AGN, and so perhaps that
  could allow the BH-bulge relation to be reached quickly. However,
  this would be surprising, because the $\sim 10 ^{11}\,M_\odot$ of
  stars in an elliptical today typically seem to have formed at higher
  redshifts \citep{daddi}.

In the context of co-evolution of AGN and their host galaxies, e.g.
the recent recently proposed by \cite{lapi}, our objects, with
$0.04<L_{FIR}/L_{BOL}<0.81$ would be in a stage when the FIR-luminous
phase is close to end (Figure 15 in \citealt{lapi}), in qualitative
agreement with our conclusion that their host galaxies are already
mostly formed.

\section{Conclusions}

We have studied a sample of X-ray-obscured QSOs at $z\sim 2$ with
strong submm emission \citep{stevens10}, much higher than most
X-ray-unobscured QSOs at similar redshifts and luminosities, which,
however, represent 85-90 per cent of the X-ray QSOs at the epoch. We
have built X-ray-to-FIR SEDs for each object and we have fitted them
using different models (a direct AGN accretion disk, a
torus-reprocessed component and a star formation component).  We
confirm that direct AGN, reprocessed AGN and SF components are needed
to correctly characterize the SEDs of our objects. We have used these
fits, together with our previous determinations of their X-ray
luminosities, to estimate the total direct and reprocessed AGN
luminosities and the luminosity associated to the star formation, as
well as other derived physical quantities.

We confirm the presence of strong FIR emission in these objects (well
above that expected from plausible AGN emission models) which we
attribute to SF at the ULIRG/HLIRG level with SFR$\sim$1000
$M_{\odot}/$y.  Their associated greybody temperature values are close
to those of Submillimeter Galaxies (SMGs). They have dust masses
around $10^9$~M$_\odot$.

We have found a just over 3-$\sigma$ significant correlation between
the SFR and the X-ray luminosity when joining our sample with that
from \cite{rovilos}. This is usually taken as evidence for joint AGN
and galaxy growth, but the differences between the techniques used to
detect and characterize SF in those two samples detract from this
otherwise exciting, interpretation.  Our objects fall in the high
luminosity end of large samples from deep surveys
\citep{rovilos,rosario}, but have similar properties to other samples
of high $z$, high FIR luminosity objects \citep{lutz}. However,
RXJ1249 stands out in all cases: it is one of the most luminous
objects known (bolometric luminosity $\sim 10^{48} L_{\odot}$) and it
has an exceptionally high torus-to-bolometric luminosity ratio for its
luminosity.

Comparing their AGN reprocessed and direct emission, we have obtained 
the ratios between the reprocessed emission from the torus and 
the total AGN bolometric luminosity $\sim 0.3-0.9$, higher than QSOs of similar
luminosities at $z<1.5$. The Eddington ratios of our objects
(0.1-0.6) are common for their redshift range, except for
RX~J1633 (our highest redshift object) which has a value of that ratio
close to unity. Overall, the bolometric corrections of our objects do
not fit well with those of other studies \citep{vasudevan,marconi04},
showing higher bolometric luminosities compared to their X-ray
luminosities.

We have found a tentative positive correlation between NH$_{ion}$ and
SFR, perhaps indicative of positive feedback between the X-ray and
UV-detected ionized outflowing gas and the interstellar medium of
their host galaxies.

The black holes powering our QSOs are very massive at their epoch,
$\sim 10^{9}-10^{10} M_{\odot}$ (measured from broad emission lines in
their optical-UV spectra). We have calculated their mass-doubling
timescale $\tau$ and the time to reach the maximum BH mass observed
locally, concluding that they can not grow much more. A further hint
in this direction comes from the high Eddington ratio of RX~J1633
which, according to \cite{kelly} should persist only for a very
brief period of time. RX~J1249 could become one of the most massive
objects known.

We do not know the masses of their host galaxies, but their black hole
masses and their high SFR lead us to conclude that they are already
very massive or they would not have enough time to reach the local
bulge-to-black-hole-mass ratio. This is also in agreement with recent
models of AGN-host galaxy co-evolution.

Direct determinations of the gas mass and of the mass of the host
galaxies our QSOs are needed to have a better grasp of the nature and
evolutionary status of these exceptional objects, and hence to
understand their role in the disputed landscape of co-evolution of
galaxies and AGN.

\section*{Acknowledgements}

The authors thank the anonymous referee for helpful suggestions to
clarify the paper. This research has made use of NASA's Astrophysics Data System
Bibliographic Services.  A.K.A thanks N. Castello for her support and
help with python. A.K.A thanks Dr. Rosario and Dr. Rovilos for sharing 
their measurements and data for the preparation of this study. 
A.K.A, F.J.C. and S.M. acknowledge financial
support from the Spanish Ministerio de Econom\'ia y Competitividad.
under project AYA2012-31447.  SM acknowledge Financial support from
the ARCHES project (7th Framework of the European Union, No. 313146).
UKIRT is operated by the Joint Astronomy Centre, Hilo, Hawaii on
behalf of the UK Science and Technology Facilities Council. Also based
on observations made with the Spitzer Space Telescope, which is
operated by the Jet Propulsion Laboratory, California Institute of
Technology, under NASA contract 1407. The James Clerk Maxwell
Telescope is operated by The Joint Astronomy Centre on behalf of the
Science and Technology Facilities Council of the United Kingdom, the
Netherlands Organisation for Scientific Research, and the National
Research Council of Canada.

This research has made use of data obtained from the SuperCOSMOS
Science Archive, prepared and hosted by the Wide Field Astronomy Unit,
Institute for Astronomy, University of Edinburgh, which is funded by
the UK Science and Technology Facilities Council. 
This publication makes use of data products from the Two
Micron All Sky Survey, which is a joint project of the University of
Massachusetts and the Infrared Processing and Analysis
Center/California Institute of Technology, funded by the National
Aeronautics and Space Administration and the National Science
Foundation. Funding for the SDSS and SDSS-II has been provided by the
Alfred P. Sloan Foundation, the Participating Institutions, the
National Science Foundation, the U.S. Department of Energy, the
National Aeronautics and Space Administration, the Japanese
Monbukagakusho, the Max Planck Society, and the Higher Education
Funding Council for England. The SDSS is managed by the Astrophysical
Research Consortium for the Participating Institutions. The
Participating Institutions are the American Museum of Natural History,
Astrophysical Institute Potsdam, University of Basel, University of
Cambridge, Case Western Reserve University, University of Chicago,
Drexel University, Fermilab, the Institute for Advanced Study, the
Japan Participation Group, Johns Hopkins University, the Joint
Institute for Nuclear Astrophysics, the Kavli Institute for Particle
Astrophysics and Cosmology, the Korean Scientist Group, the Chinese
Academy of Sciences (LAMOST), Los Alamos National Laboratory, the
Max-Planck-Institute for Astronomy (MPIA), the Max-Planck-Institute
for Astrophysics (MPA), New Mexico State University, Ohio State
University, University of Pittsburgh, University of Portsmouth,
Princeton University, the United States Naval Observatory, and the
University of Washington.
\noindent Based on observations made with the William Herschel
Telescope and its service programme-operated by the Isaac Newton
Group, installed in the Spanish Observatorio del Roque de los
Muchachos of the Instituto de Astrofsica de Canarias, in the island of
La Palma. This work is based on observations obtained with Herschel
Space Telescope, an ESA science mission with instruments and
contributions directly funded by ESA Member States. Based on data from
the Wide-field Infrared Survey Explorer, which is a joint project of
the University of California, Los Angeles, and the Jet Propulsion
Laboratory/ California Institute of Technology, funded by the National
Aeronautics and Space Administration.

This research has made use TOPCAT software
(http://www.starlink.ac.uk/topcat/) and its tools.

\section*{Appendix}
We now discuss briefly the fit results for each source:
\begin{itemize}

\item RX~J0057: We have chosen the first torus model family from
  \cite{roseboom} (11 members), since the other fits gave $\chi^2$
  values at least 40 per cent higher.
 
\item RX~J0941: We have selected the first torus model family from
  \cite{roseboom} and the empirical template from \cite{rowan} (22
  members). The other torus models from \cite{roseboom} have $\chi^2$
  values at least 30 per cent higher.
 
\item RX~J1218: Contrary to the other sources, the best-fit models
  were not visually good, since the $\chi^2$ value was dominated by
  the fits to the optical-to-MIR region (where the error bars are
  smallest), leaving the SF bump badly represented. For each
  disk+torus+SF combination we fixed the model parameters to their
  values best fitting the overall SED, and re-calculated the $\chi^2$
  value restricted to the rest-frame band 36-190~$\mu$m (5 points). We
  found that this restricted $\chi^2$ had a clear best-fit peak with
  $\chi^2\leq 12$, including 20 disk+torus+SF combinations, with the
  rest of the combinations extending in a tail towards higher
  restricted $\chi^2$ values. We have chosen these 20 combinations as our
  best-fit ``family''. Their torus components include both the first
  and second torus models of \cite{roseboom}, as well as the empirical
  template from \cite{rowan}.
 
\item RX~J1249: This case is very similar to RX~J0057: we have selected
  the first torus model family from \cite{roseboom}, for which the $\chi^2$
  values were at least 40 per cent lower.
 
\item RX~J1633: The third torus model from \cite{roseboom} produced a
  well-grouped family (11 members) of best-fits with low $\chi^2$
  values. There were a few scattered best-fits with lower $\chi^2$
  using other models but they showed systematic residuals around
  rest-frame 30-60~$\mu$m, with SF components taking over the torus
  contribution in that band. We have chosen that family as our
  preferred fit.

\end{itemize}

From the results, see Figure~\ref{figAllFits}, above, it is clear that
the first torus model of \cite{roseboom} is preferred by two
of our sources (RX~J0057 and RX~J1249), and acceptable for another two
(RX~J0941 and RX~J1218), for which the empirical model of \cite{rowan}
is also admitted. The second torus model of \cite{roseboom} only
appears among the best-fits in one case (RX~J1218). Finally, the third
torus model in \cite{roseboom} is the best representation for RX~J1633
\appendix

 \bsp

\label{lastpage}

\end{document}